\documentclass[a4paper,12pt, epsfig]{article}
\def\letter{0}\def\pr{0}
\pdfoutput=1 %per jhep
\usepackage{epsfig}
\usepackage{epstopdf}
\usepackage{graphicx}
\usepackage{ifthen}
\usepackage{slashed}

\usepackage{amsmath}
\usepackage[psamsfonts]{amssymb}
\usepackage{euscript}

\usepackage{latexsym}
\usepackage[arrow,matrix,curve]{xy}

\newskip\humongous \humongous=0pt plus 1000pt minus 1000pt

\newif\ifdtup

\def\,{\hspace{-.1cm}}
\def\hsp{,\hspace{.7cm}}

\def\fc#1#2 {\frac{n}{q}#1\frac{n}{q}#2}

\renewcommand{\theequation}{\arabic{section}.\arabic{equation}}
\renewcommand{\(}{\begin{equation}}
\renewcommand{\)}{end{equation} \vspace{-.05in}\linebreak}

\newcounter{saveeqn}
\newcounter{savealpheqn}

\newcommand{\alpheqn}{\setcounter{saveeqn}{\value{equation}}%
  \stepcounter{saveeqn}\setcounter{equation}{0}%
  \renewcommand{\theequation}{\mbox{\arabic{section}.\arabic{saveeqn}
\alph{equation}}}
  \renewcommand{\)}{\end{equation}}}
\def\part#1{\frac{\partial}{\partial{#1}}}%
\def\group#1{\refstepcounter{equation}\setcounter{saveeqn}
 {\value{equation}}%
  \label{#1}\setcounter{equation}{0}%
\renewcommand{\theequation}{\mbox{\arabic{section}.\arabic{saveeqn}
\alph{equation}}}
  \renewcommand{\)}{\end{equation}}}
\newcommand{\reseteqn}{\setcounter{equation}{\value{saveeqn}}%
  \renewcommand{\theequation}{\arabic{section}.\arabic{equation}}%
  \renewcommand{\)}{\end{equation}}}

\newcommand{\aalpheqn}{\setcounter{saveeqn}{\value{equation}}%
  \stepcounter{saveeqn}\setcounter{equation}{0}%
  \renewcommand{\theequation}{\mbox{
        \Alph{subsection}.\arabic{saveeqn}\alph{equation}}}
   \renewcommand{\)}{\end{equation}}}
\newcommand{\areseteqn}{\setcounter{equation}{\value{saveeqn}}%
  \renewcommand{\theequation}{\Alph{subsection}.\arabic{equation}}%
  \renewcommand{\)}{\end{equation}}}

\renewcommand{\=}{\hspace{-.03in}=\hspace{-.02in}}
\renewcommand{\thefootnote}{\alph{footnote}}
\renewcommand{\(}{\begin{equation}}
\renewcommand{\)}{\end{equation}}
\newcommand{\ba}{\begin{eqnarray}}
\newcommand{\ea}{\end{eqnarray}}
\newcommand{\cbp}{\mathop{\vtop{\ialign{##\crcr
   $\hfil\displaystyle{}\hfil$\crcr\noalign{\kern-13pt\nointerlineskip}
   \BIG{)}\hskip0pt\crcr\noalign{\kern3pt}}}}}
\newcommand{\pa}{\mathop{\vtop{\ialign{##\crcr

$\hfil\displaystyle{\oplus}\hfil$\crcr\noalign{\kern+1pt\nointerlineskip
}
   \hspace{.08in}$^{\alpha=0}$\hskip6pt\crcr\noalign{\kern3pt}}}}}
\renewcommand{\hsp}{,\hspace{.3in}}
\newcommand{\p}{^\prime}

\catcode`\@=11
\def\vereq#1#2{\lower3pt\vbox{\baselineskip1.5pt \lineskip1.5pt
\ialign{$\m@th#1\hfill##\hfil$\crcr#2\crcr\sim\crcr}}}
\catcode`\@=12

\renewcommand{\(}{\begin{equation}}
\renewcommand{\)}{\end{equation}}

\def\pina#1#2{\int \frac{d^{#2}{#1}}{(2\pi)^{#2}}}

\newcommand{\beas}{\begin{eqnarray*}}
\newcommand{\eeas}{\end{eqnarray*}}

\newcommand{\bquo}{\begin{quote}}
\newcommand{\enqu}{\end{quote}}

\newcommand{\beq}{\begin{equation}}
\newcommand{\eeq}{\end{equation}}
\newcommand{\bea}{\begin{eqnarray}}
\newcommand{\eea}{\end{eqnarray}}

\newskip\humongous \humongous=0pt plus 1000pt minus 1000pt

\newif\ifdtup

\catcode`\@=11

\ifthenelse{\equal{\letter}{0}}{ %se lettere=0, commenta questo

\@addtoreset{equation}{section}
\def\theequation{\arabic{section}.\arabic{equation}}

\def\@normalsize{\@setsize\normalsize{15pt}\xiipt\@xiipt
\abovedisplayskip 14pt plus3pt minus3pt%
\belowdisplayskip \abovedisplayskip
\abovedisplayshortskip \z@ plus3pt%
\belowdisplayshortskip 7pt plus3.5pt minus0pt}

\def\small{\@setsize\small{13.6pt}\xipt\@xipt
\abovedisplayskip 13pt plus3pt minus3pt%
\belowdisplayskip \abovedisplayskip
\abovedisplayshortskip \z@ plus3pt%
\belowdisplayshortskip 7pt plus3.5pt minus0pt
\def\@listi{\parsep 4.5pt plus 2pt minus 1pt
      \itemsep \parsep
      \topsep 9pt plus 3pt minus 3pt}}

\relax

\def\section{\@startsection{section}{1}{\z@}{3.5ex plus 1ex minus  .2ex}{2.3ex plus .2ex}{\large\bf}}

\def\thesection{\arabic{section}}
\def\thesubsection{\arabic{section}.\arabic{subsection}}

\def\appendix{\setcounter{section}{0}
 \def\thesection{Appendix \Alph{section}}
 \def\thesubsection{\Alph{section}.\arabic{subsection}}
 \def\theequation{\Alph{section}.\arabic{equation}}}
\renewcommand{\theequation}{\arabic{section}.\arabic{equation}}

}{
\renewcommand{\theequation}{\arabic{equation}}

} %se letter=0, commenta questo

\begin{document}
% ========================================================================
\def\thefootnote{\fnsymbol{footnote}}
\def\thetitle{Where Neutrino Decoherence Lies}
\def\auttwo{Jarah Evslin}
\def\autone{Emilio Ciuffoli}
\def\affa{Institute of Modern Physics, NanChangLu 509, Lanzhou 730000, China}
\def\affb{University of the Chinese Academy of Sciences, YuQuanLu 19A, Beijing 100049, China}

\title{Titolo}

\ifthenelse{\equal{\pr}{1}}{
\title{\thetitle}
\author{\autone}
\author{\auttwo}
\affiliation {\affa}
\affiliation {\affb}
%pr e uno

}{

\begin{center}
{\large {\bf \thetitle}}

\bigskip

\bigskip

%\catcode`@=11

{\large \noindent  \autone{${}^{1}$} and \auttwo{${}^{1,2}$}}

%{\large \noindent  \autone{${}^{1,2}$} \footnote{jarah@impcas.ac.cn} and \auttwo{${}^{1,2}$} \footnote{guohengyuan@impcas.ac.cn}}

\vskip.7cm

1) \affa\\
2) \affb\\
%3) \affc\\

\end{center}

}

\begin{abstract}
\noindent
Recently, several studies of neutrino oscillations in the vacuum have not found the decoherence long expected from the separation of wave packets of neutrinos in different mass eigenstates.  We show that such decoherence will, on the other hand, be present in a treatment including any mechanism which leads to a dependence of the final state on both the neutrino's emission and absorption time.  Our demonstration is in the 3+1d $V-A$ model, however the details of this model lead to an overall factor which does not affect our conclusions.  This allows us to consider a simpler model.  There we show that if the positions of the final state particles are measured, or equivalently entangled with the environment, then decoherence will damp neutrino oscillations.  We also show that wave packet spreading can cause the decoherence to eventually saturate, without completely suppressing the oscillations.

%We show in a simple example that the strong time-dependence required for significant decoherence does not imply that the energy resolution of the detector is insufficient to detect oscillations and that decoherence is observable if the final states are described by localized wave packets, even if no external constraint is placed on the neutrino creation and detection time. We also show that wave packet spreading can cause the decoherence to eventually saturate, without completely suppressing the oscillations.%We also show that if terms beyond the first-order expansion of the neutrino energy are considered, the decoherence can saturate after a certain point, without suppressing completely the oscillations.

\end{abstract}

% \vfill
%
% \end{titlepage}
\setcounter{footnote}{0}
\renewcommand{\thefootnote}{\arabic{footnote}}

\ifthenelse{\equal{\pr}{1}}
{
\maketitle
}{}

\section{Introduction}
Neutrinos are described by wave packets.  Lighter neutrinos in general correspond to faster wave packets which eventually are expected to pull ahead of wave packets corresponding to heavier neutrinos, ruining their coherence and thus the observed pattern of neutrino oscillations.  This old expectation has been confirmed in quantum mechanics \cite{nuss76,kayser81,rich93} and in external wave packet models \cite{giunti93,beuthe}.  However, Nature is described by quantum field theory with dynamical fields.  In the full, dynamical quantum field theory, in a vacuum, Refs.~\cite{mcgreevy,grimus,misurapprox} found no such decoherence.  Furthermore, Ref.~\cite{mcdonald} argued that decoherence, in a vacuum or coupled to an environment, would anyway be unobservable as it can appear only when the detector's energy resolution is too poor to observe neutrino oscillations.  This state of affairs has led some to believe that neutrino quantum decoherence cannot be observed even in principle, even when the system is coupled to an environment, despite many calculations \cite{coelho17,gomes20} in simplified models which suggest the contrary.

In the present note, we examine how the arguments in the above vacuum studies fail in the presence of an environment.   The decoherence is a result of the fact that the final state of the environment depends on the spacetime region in which the neutrino is created and absorbed, which in turn depend on the neutrino mass eigenstate.   We will see that this statement is quite independent of the details of the model and so, in a simplified model, we provide a counter example to one step in the argument of Ref.~\cite{mcdonald}. 

%We also show that, if the time of production and detection of the neutrino is constrained, decoherence would arise; this can be achieved by turning on the interactions only for a limited window of time or by considering localized final state, which would constrain the region of production of the neutrino for kinematic reasons.

Often, in order to obtain analytical results, energies are expanded up to first order in momentum.  At this order the entire wave packet moves rigidly, at a constant velocity. We have considered the second order expansion in a simplified model.  We found that the dynamical evolution of the neutrino wave packet can affect the decoherence: at first the oscillation amplitude decreases, however at some point the decoherence can saturate, leading to a regime where the partially dampened oscillations are not further damped.

\subsection{Three Sources of Decoherence}\label{subSecThreeSources}
In a quantum theory, the probability of a given set of final states $\{|i\rangle\}$ given a known initial state is calculated via two sums.  First, one sums the amplitudes for all processes yielding a given final state $|i\rangle$.  This sum of complex numbers is called the {\it{coherent sum}}, and it results in the total amplitude for each final state.  Next, one takes the norm squared of each total amplitude and sums these over the set of final states of interest $\{|i\rangle\}$.  This sum of real numbers is called the {\it{incoherent sum}}.   

While each term in a coherent sum is an amplitude with the same final state, the intermediate states are in general different.  For example, an internal neutrino line may correspond to different mass eigenstates in different terms in the sum, so long as these different mass eigenstates lead to the same final state.  When this is the case, the coherent sum of different mass eigenstate processes leads to neutrino oscillations.  On the other hand, oscillations will be washed out by quantum decoherence if distinct mass eigenstates at intermediate steps lead to distinct final states\footnote{This is not to be confused with classical decoherence, which was observed more than 20 years ago and arises from incoherent sums, as a result of distinct final states which cannot be distinguished due to limited energy resolution or a poor identification of the source location.  For a recent review, see Ref.~\cite{cheng22}.  We also do not consider dissipative effects caused by new physics \cite{lisi00,benatti00}.}.

Do distinct neutrino mass eigenstates always lead to the same final state?  If the neutrino propagates over a long baseline, then it will nearly be on shell and then the relevant kinematic constraints in many experiments of interest imply that the heavier mass eigenstates will travel more slowly.  This means that the space-time point at which the neutrino is created or absorbed will, in such cases, differ for different mass eigenstates\footnote{The wave packet description of neutrinos arises from the same mechanism.  The neutrino trajectories which lead to the same final state sum coherently and so can be treated as a coherent wave packet.  A concrete connection between the finiteness of the spacetime production region and the neutrino wave packet in quantum field theory was provided in Refs.~\cite{ak10,as10}.}.  As a result, {\bf{if the final state is sufficiently sensitive to the neutrino emission and absorption points, then the distinct mass eigenstates will lead to distinct final states and will be incoherently summed, leading to a damping in the oscillation amplitude.}}

Why might the final states depend on the space-time locations of the neutrino emission and absorption?  There are three physical mechanisms for such a dependence.  First, the macroscopic observables of the experiment inevitably place some constraint on the neutrino emission and absorption.  For example, in a terrestrial neutrino experiment, emission is generally spatially localized at a radioactive source, at a target station or in a decay tube.  Absorption is spatially localized in a detector.  Emission is temporally localized as a result of the finite lifetime of the source or the beam structure.  Detection is temporally localized by the time resolution of the detector.  As was shown in Ref.~\cite{grimus}, such macroscopic constraints are far too weak to lead to observable decoherence and we will not consider them further.

The second mechanism is that the processes in which the neutrino is emitted and absorbed will inevitably affect the environment, so long as they do not occur in a vacuum.  The final state of the environment therefore will depend on the locations of both the neutrino emission and also its absorption.  

Third, lepton number conservation together with the light mass of the neutrino implies that neutrinos are generally emitted together with charged leptons, and when they are absorbed charged leptons are generally created.  These charged leptons survive to the final state in many experimental setups of interest and so these final states will generally be sensitive to the space-time locations where the charged leptons are created\footnote{The entanglement of these charged leptons with the neutrinos has been stressed in Ref.~\cite{cgl}.}, which are precisely the locations where the neutrino is emitted and absorbed.  Thus, for each final state, one expects decoherence.  If this decoherence survives when all experimentally indistinguished final states are summed, it will provide an irreducible contribution to the decoherence.  This irreducible source of decoherence is present in the vacuum and can be calculated precisely using the Standard Model, or low energy approximations to it such as the $V-A$ model.

For simplicity, following Ref.~\cite{kw98} we will consider only the environmental dependence on the time of the neutrino emission and absorption, as the position is anyway constrained by the dimensions of the source and detector.  However, we expect the environmental dependence on the position of emission and absorption to dominate over the position constraint arising from the source and detector sizes, and so in a quantitative treatment it should be included as well.  Unlike the irreducible decoherence, this environmental decoherence depends strongly on the details of the environmental interaction.  We will parametrize this dependence with a parameter $\sigma$, which is the size of the time window that leads to the same final state.  We will use the same $\sigma$ for both emission and absorption.  Although this is certainly not the case in practice, the generalization to two different values of $\sigma$ is obvious.

Using a simplified model we will also show that the space-time localization of the products of the decay will impose a similar constraint on the neutrino production time, allowing decoherence to emerge. Such a process is very similar to the third mechanism described, albeit in a rather simplified scenario. A more complete treatment will be provided in a companion paper.

%%%% END OF MODIFIED VERSION

\subsection{Outline}
In Sec.~\ref{secGrimusGen} we consider the 3+1d $V-A$ model, repeating the calculation of Ref.~\cite{grimus} but including an environment which is sensitive to the neutrino production and absorption time.  We find that, when the environment is considered, there is indeed decoherence.   We observe that this decoherence is independent of the details of the model, which affect only the normalization of the oscillation probability.  This observation allows us, in Sec.~\ref{secModel} to present a simplified model, containing only scalar fields in 1+1d, that will be used to compute the shape and dimension of the neutrino wave packet as a function of $t$.  We will see that, even if the uncertainty on the momentum of the source particle leads to a spread in the neutrino momentum, their widths are not the same: the dimension of the neutrino wave packet is suppressed by a factor of $v_I-v_F\simeq q/M$, where $q$ is the neutrino momentum and $M$ the source particle mass.  This mismatch addresses the objection of Ref.~\cite{mcdonald}.  In Sec. \ref{secAmp} we use this simplified model to compute the transition amplitude, showing that if the final states are localized as well, decoherence can emerge. We will also show that, if the time-evolution of the neutrino wave packet is taken into account, decoherence will not completely erase the oscillations.

\section{Decoherence from Environmental Interactions}
\label{secGrimusGen}

\subsection{Defining the Amplitude in the $V-A$ Model} \label{secGrimus}

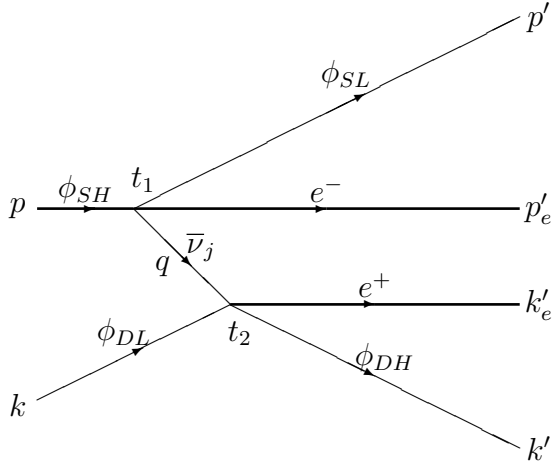
\begin{figure}
\setlength{\unitlength}{.5in}
\begin{picture}(3,4)(-6,-1.5)
\linethickness{1pt}
%\put(-.3,-1){\makebox(0,0){{\bf{Process A}}}}
\put(-2,1){\vector(1,0){.6}}
\put(-1.5,1){\line(1,0){.5}}
\put(-1.5,1.2){\makebox(0,0){$\phi_{SH}$}}
\put(-2.2,1.0){\makebox(0,0){$p$}}

\put(-.9,1.3){\makebox(0,0){$t_1$}}

\put(-1,1){\vector(2,1){2.4}}
\put(1,2){\line(2,1){2}}
\put(1.2,2.4){\makebox(0,0){$\phi_{SL}$}}
\put(3.2,3){\makebox(0,0){$p\p$}}

\put(-1,1){\vector(1,0){2}}
\put(1,1){\line(1,0){2}}
\put(1,1.2){\makebox(0,0){$e^-$}}
\put(3.2,1){\makebox(0,0){$p_e\p$}}

\put(-1,1){\vector(1,-1){.6}}
\put(-.5,.5){\line(1,-1){.5}}
\put(-.3,.6){\makebox(0,0){$\overline{\nu}_j$}}
\put(-.7,.4){\makebox(0,0){$q$}}

\put(0,0){\vector(2,-1){1.5}}
\put(1.5,-.75){\line(2,-1){1.5}}
\put(1.6,-.55){\makebox(0,0){$\phi_{DH}$}}
\put(3.2,-1.5){\makebox(0,0){$k\p$}}

\put(0,0){\vector(1,0){1.5}}
\put(1.5,0){\line(1,0){1.5}}
\put(1.5,.2){\makebox(0,0){$e^+$}}
\put(3.2,0){\makebox(0,0){$k_e\p$}}

\put(.1,-.3){\makebox(0,0){$t_2$}}

\put(-2,-1){\vector(2,1){1.1}}
\put(-1,-.5){\line(2,1){1}}
\put(-1.1,-.29){\makebox(0,0){$\phi_{DL}$}}
\put(-2.2,-1.05){\makebox(0,0){$k$}}
\end{picture}

\caption{Time flows to the right.  A heavy source particle $\beta$-decays into a light source particle, an electron and a antineutrino.  The antineutrino is absorbed by the light detector particle, which becomes a heavy detector particle and emits a positron.  The arrows indicate the signs of the momenta.}
\label{procfig}
\end{figure}

In this section, we will consider a setup similar to that of Ref.~\cite{grimus}.  It is an initial value problem in quantum field theory, beginning at time $t_S$ and ending at time $t_D$.  The initial state is fixed at time $t_S$ and we will calculate the probability of various final states at time $t_D$.  We describe the emission of an antineutrino in the $\beta$-decay process
\beq
\phi_{SH}(p)\longrightarrow \phi_{SL}(p\p)+\overline{\nu}_j(q)+e^-(p\p_e)
\eeq
at time $t_1$ followed by its absorption via inverse $\beta$-decay
\beq
\phi_{DL}(k)+\overline{\nu}_j(q)\longrightarrow \phi_{DH}(k\p)+e^+(k\p_e)
\eeq
at time $t_2$, all in the $V-A$ model.  The total process is drawn in Fig.~\ref{procfig}.

While Ref.~\cite{grimus} considers this process in a vacuum, we do not.  Instead, we consider the system coupled to an environment in the following approximation.  The environment does not affect the system, however the final state of the environment depends on $t_1$ and $t_2$.  More precisely, the final state of the environment is described by two integer quantum numbers $|IJ\rangle$.  If the neutrino is emitted at time $t_1$ then the amplitude for the first quantum number to have a fixed value $I$ is
\beq
\frac{e^{-\left(t_1-\tau_{S,I}\right)^2/4\sigma^2}}{(2\pi)^{1/4}\sqrt{\sigma}}\hsp
\tau_{S,I}=t_S+I \sigma \label{pamp}
\eeq
where $\sigma$ is a real parameter, the only parameter in our model.  Similarly, if the neutrino is absorbed at time $t_2$ then the amplitude for a fixed second quantum number $J$ is
\beq
\frac{e^{-\left(t_2-\tau_{D,J}\right)^2/4\sigma^2}}{(2\pi)^{1/4}\sqrt{\sigma}}\hsp
\tau_{D,J}=t_D+J \sigma. \label{damp}
\eeq

The initial value problem is then completely determined by the initial condition.  The initial condition is that the heavy source and light detector particles are described by the wave functions $\psi_S$ and $\psi_D$, so that the initial state is
\beq
\pina p3 \psi_S(\vec{p})\pina k3 \psi_D(\vec{k})|\vec{p},\vec{k}\rangle.
\eeq

In signature (+,-,-,-), we may then write the amplitude for the final state with momenta $p\p,\ p\p_e,\ k\p$ and $k\p_e$ and environmental final state $|IJ\rangle$.  It is
\bea
\mathcal{A}_{I,J}(p\p,p\p_e,k\p,k\p_e)&=&-i\sum_{j=1}^3\int d^4 x_1\int d^4 x_2 \pina q4 e^{-iq\cdot (x_2-x_1)}\\
&&\times\pina p3 \psi_S(\vec{p})e^{-ip\cdot(x_1-x_S)}\frac{e^{-\left(t_1-\tau_{S,I}\right)^2/4\sigma^2}}{(2\pi)^{1/4}\sqrt{\sigma}}e^{i(p\p+p\p_e)\cdot x_1}J_{S}^\lambda(\vec{p}\p,\vec{p})\nonumber\\
&&\times \pina k3 \psi_D(\vec{k})e^{-ik\cdot(x_2-x_D)}
\frac{e^{-\left(t_2-\tau_{D,J}\right)^2/4\sigma^2}}{(2\pi)^{1/4}\sqrt{\sigma}}e^{i(k\p+k\p_e)\cdot x_2}J_{D}^\rho(\vec{k}\p,\vec{k})\nonumber\\
&&\times \left|U_{ej}\right|^2 \overline{u}_e(p\p_e)\gamma_\lambda(1-\gamma_5)\frac{\slashed{q}+m_j}{q^2-m_j^2+i\epsilon}\gamma_\rho (1-\gamma_5)v_e(k\p_e)\nonumber
\eea
where $J$ is the usual hadronic form factor \cite{grimus}
\bea
\frac{G_F}{\sqrt{2}}\langle \phi_{SL}(p\p)|\overline{u}(x)\gamma^\lambda(1-\gamma_5)U_{\rm{CKM}}d(x)|\phi_{SH}(p)\rangle&=&J_{S}^\lambda(p\p,p)e^{i(p\p-p)\cdot x}\\
\frac{G_F}{\sqrt{2}}\langle \phi_{DH}(k\p)|\left(\overline{u}(x)\gamma^\rho(1-\gamma_5)U_{\rm{CKM}}d(x)|\phi_{SH}(p)\right)^\dag|\phi_{DL}(k)\rangle&=&J_{D}^\rho(k\p,k)e^{i(k\p-k)\cdot x}.\nonumber
\eea
Here $m_j$ is the mass of the $j$th neutrino mass eigenstate, which is related to the electron via the element $U_{ej}$ of the $PMNS$ matrix.

Unlike Ref.~\cite{grimus}, we consider neither the finite lifetime of the source nor the data taking time, as the author of that paper showed that neither has a significant effect on decoherence.  Our scale $\sigma$, in order to be phenomenologically relevant, must be much shorter than both of these scales. 

\subsection{Evaluating the Amplitude}
\label{secGrimusAmp}

Integrate over $\vec{x}_1$ and $\vec{x}_2$
\bea
\mathcal{A}_{I,J}(p\p,p\p_e,k\p,k\p_e)&=&-i\sum_{j=1}^3\int dt_1\int dt_2 \pina q4 e^{-iq_0 (t_2-t_1)}\\
&&\times\pina p3 (2\pi)^3\delta^3(\vec{p}-\vec{q}-\vec{p}\p-\vec{p}\p_e)\psi_S(\vec{p})e^{-ip_0(t_1-t_S)}e^{-i\vec{p}\cdot\vec{x}_S}\nonumber\\
&&\times\frac{e^{-\left(t_1-\tau_{S,I}\right)^2/4\sigma^2}}{(2\pi)^{1/4}\sqrt{\sigma}}e^{i(p\p_0+p\p_{e,0}) t_1}J_{S}^\lambda(\vec{p}\p,\vec{p})\nonumber\\
&&\times \pina k3 (2\pi)^3\delta^3(\vec{k}+\vec{q}-\vec{k}\p-\vec{k}\p_e)\psi_D(\vec{k})e^{-ik_0(t_2-t_D)}e^{-i\vec{k}\cdot\vec{x}_D}\nonumber\\
&&\times\frac{e^{-\left(t_2-\tau_{D,J}\right)^2/4\sigma^2}}{(2\pi)^{1/4}\sqrt{\sigma}}e^{i(k\p_0+k\p_{e,0}) t_2}J_{D}^\rho(\vec{k}\p,\vec{k})\nonumber\\
&&\times \left|U_{ej}\right|^2 \overline{u}_e(p\p_e)\gamma_\lambda(1-\gamma_5)\frac{\slashed{q}+m_j}{q^2-m_j^2+i\epsilon}\gamma_\rho (1-\gamma_5)v_e(k\p_e).\nonumber
\eea
Define
\beq
E_S=\sqrt{m_N^2+\left|\vec{q}+\vec{p}\p+\vec{p}\p_e\right|^2}\hsp
E_D=\sqrt{m_P^2+\left|-\vec{q}+\vec{k}\p+\vec{k}\p_e\right|^2}
\eeq
and then integrate over $p$ and $k$
\bea
\mathcal{A}_{I,J}(p\p,p\p_e,k\p,k\p_e)&=&-i\sum_{j=1}^3\int dt_1\int dt_2 \pina q4 e^{-iq_0 (t_2-t_1)}\\
&&\times\psi_S(\vec{q}+\vec{p}\p+\vec{p}\p_e)e^{-iE_S(t_1-t_S)}e^{-i(\vec{q}+\vec{p}\p+\vec{p}\p_e)\cdot\vec{x}_S}\nonumber\\
&&\times\frac{e^{-\left(t_1-\tau_{S,I}\right)^2/4\sigma^2}}{(2\pi)^{1/4}\sqrt{\sigma}}e^{i(p\p_0+p\p_{e,0}) t_1}J_{S}^\lambda(\vec{p}\p,\vec{q}+\vec{p}\p+\vec{p}\p_e)\nonumber\\
&&\times \psi_D(-\vec{q}+\vec{k}\p+\vec{k}\p_e)e^{-iE_D(t_2-t_D)}e^{-i(-\vec{q}+\vec{k}\p+\vec{k}\p_e)\cdot\vec{x}_D}\nonumber\\
&&\times\frac{e^{-\left(t_2-\tau_{D,J}\right)^2/4\sigma^2}}{(2\pi)^{1/4}\sqrt{\sigma}}e^{i(k\p_0+k\p_{e,0}) t_2}J_{D}^\rho(\vec{k}\p,-\vec{q}+\vec{k}\p+\vec{k}\p_e)\nonumber\\
&&\times \left|U_{ej}\right|^2 \overline{u}_e(p\p_e)\gamma_\lambda(1-\gamma_5)\frac{\slashed{q}+m_j}{q^2-m_j^2+i\epsilon}\gamma_\rho (1-\gamma_5)v_e(k\p_e).\nonumber
\eea
Now integrate over $t_1$ and $t_2$
\bea
\mathcal{A}_{I,J}(p\p,p\p_e,k\p,k\p_e)&=&-4i\sigma\sqrt{2\pi}e^{-i(\vec{p}\p+\vec{p}\p_e)\cdot\vec{x}_S}e^{-i(\vec{k}\p+\vec{k}\p_e)\cdot\vec{x}_D}e^{i(p\p_0+p\p_{e,0}) \tau_{S,I}}e^{i(k\p_0+k\p_{e,0})\tau_{D,J}}\label{tint}\\
&&\times\sum_{j=1}^3\pina q4 e^{-iq_0 (\tau_{D,J}-\tau_{S,I})}\psi_S(\vec{q}+\vec{p}\p+\vec{p}\p_e)e^{-iE_S(\tau_{S,I}-t_S)}\nonumber\\
&&\times e^{- \sigma^2\left(E_S-q_0-p\p_0-p\p_{e,0}\right)^2}J_{S}^\lambda(\vec{p}\p,\vec{q}+\vec{p}\p+\vec{p}\p_e)\nonumber\\
&&\times \psi_D(-\vec{q}+\vec{k}\p+\vec{k}\p_e)e^{-iE_D(\tau_{D,J}-t_D)}e^{i\vec{q}\cdot(\vec{x}_D-\vec{x}_S)}\nonumber\\
&&\times e^{-\sigma^2\left(E_D+q_0-k\p_0-k\p_{e,0}\right)^2}J_{D}^\rho(\vec{k}\p,-\vec{q}+\vec{k}\p+\vec{k}\p_e)\nonumber\\
&&\times \left|U_{ej}\right|^2 \overline{u}_e(p\p_e)\gamma_\lambda(1-\gamma_5)\frac{\slashed{q}+m_j}{q^2-m_j^2+i\epsilon}\gamma_\rho (1-\gamma_5)v_e(k\p_e).\nonumber
\eea

We define
\beq
 \vec{x}_D-\vec{x}_S=\vec{L}=L \hat{L}
\eeq
where $\hat{L}$ is a unit vector.  At large $L$, using the Grimus-Stockinger Theorem \cite{gs}
\beq
\stackrel{lim}{{}_{L\rightarrow\infty}}\pina q 3 \frac{\Phi(\vec{q})}{q_0^2-m_j^2-|\vec{q}|^2+i\epsilon}e^{i\vec{q}\cdot \vec{L}}=-\frac{1}{4\pi L} \Phi(\sqrt{q_0^2-m_j^2}\hat{L})e^{i\sqrt{q_0^2-m_j^2}L}.
\eeq
%Anyway, there is no decoherence at $\sigma=\infty$ because the neutrinos all have plenty of time to arrive.  So 
Let us try the $\vec{q}$ integral directly on (\ref{tint})
\bea
\mathcal{A}_{I,J}(p\p,p\p_e,k\p,k\p_e)&=&-4i\sigma\sqrt{2\pi}e^{-i(\vec{p}\p+\vec{p}\p_e)\cdot\vec{x}_S}e^{-i(\vec{k}\p+\vec{k}\p_e)\cdot\vec{x}_D}e^{i(p\p_0+p\p_{e,0}) \tau_{S,I}}e^{i(k\p_0+k\p_{e,0})\tau_{D,J}}\\
&&\times\sum_{j=1}^3\int\frac{dq_0}{2\pi} e^{-iq_0 (\tau_{D,J}-\tau_{S,I})}\psi_S(\sqrt{q_0^2-m_j^2}\hat{L}+\vec{p}\p+\vec{p}\p_e)e^{-iE_S(\tau_{S,I}-t_S)}\nonumber\\
&&\times e^{- \sigma^2\left(E^j_S-q_0-p\p_0-p\p_{e,0}\right)^2}J_{S}^\lambda(\vec{p}\p,\sqrt{q_0^2-m_j^2}\hat{L}+\vec{p}\p+\vec{p}\p_e)\nonumber\\
&&\times \psi_D(-\sqrt{q_0^2-m_j^2}\hat{L}+\vec{k}\p+\vec{k}\p_e)e^{-iE_D(\tau_{D,J}-t_D)}e^{i\sqrt{q_0^2-m_j^2}L}\nonumber\\
&&\times e^{-\sigma^2\left(E^j_D+q_0-k\p_0-k\p_{e,0}\right)^2}J_{D}^\rho(\vec{k}\p,-\sqrt{q_0^2-m_j^2}\hat{L}+\vec{k}\p+\vec{k}\p_e)\nonumber\\
&&\times \left|U_{ej}\right|^2 \overline{u}_e(p\p_e)\gamma_\lambda(1-\gamma_5)\left(\slashed{q}|_{\vec{q}=\sqrt{q_0^2-m_j^2}\hat{L}}+m_j\right)\gamma_\rho (1-\gamma_5)v_e(k\p_e)\nonumber
\eea
where it is understood that $E^j_S$ and $E^j_D$ are evaluated at $\vec{q}=\sqrt{q_0^2-m_j^2}\hat{L}$
\beq
E^j_S=\sqrt{m_N^2+\left|\sqrt{q_0^2-m_j^2}\hat{L}+\vec{p}\p+\vec{p}\p_e\right|^2}\hsp
E^j_D=\sqrt{m_P^2+\left|-\sqrt{q_0^2-m_j^2}\hat{L}+\vec{k}\p+\vec{k}\p_e\right|^2}.
\eeq
In particular at $I=J=0$, letting
\beq
T=t_D-t_S
\eeq
we find
\bea
\mathcal{A}_{00}(p\p,p\p_e,k\p,k\p_e)&=&-4i\sigma\sqrt{2\pi}e^{-i(\vec{p}\p+\vec{p}\p_e)\cdot\vec{x}_S}e^{-i(\vec{k}\p+\vec{k}\p_e)\cdot\vec{x}_D}e^{i(p\p_0+p\p_{e,0}) t_S}e^{i(k\p_0+k\p_{e,0})t_D}\\
&&\times\sum_{j=1}^3\int\frac{dq_0}{2\pi} e^{-iq_0 T}e^{i\sqrt{q_0^2-m_j^2}L}\nonumber\\
&&\times e^{- \sigma^2\left(E^j_S-q_0-p\p_0-p\p_{e,0}\right)^2} e^{-\sigma^2\left(E^j_D+q_0-k\p_0-k\p_{e,0}\right)^2}\nonumber\\
&&\times\psi_S\left(\sqrt{q_0^2-m_j^2}\hat{L}+\vec{p}\p+\vec{p}\p_e\right) \psi_D\left(-\sqrt{q_0^2-m_j^2}\hat{L}+\vec{k}\p+\vec{k}\p_e\right)\nonumber\\
&&\times J_{S}^\lambda\left(\vec{p}\p,\sqrt{q_0^2-m_j^2}\hat{L}+\vec{p}\p+\vec{p}\p_e\right)J_{D}^\rho\left(\vec{k}\p,-\sqrt{q_0^2-m_j^2}\hat{L}+\vec{k}\p+\vec{k}\p_e\right)\nonumber\\
&&\times \left|U_{ej}\right|^2 \overline{u}_e(p\p_e)\gamma_\lambda(1-\gamma_5)\left(\slashed{q}|_{\vec{q}=\sqrt{q_0^2-m_j^2}\hat{L}}+m_j\right)\gamma_\rho (1-\gamma_5)v_e(k\p_e).\nonumber
\eea
Let us ignore the flavor dependence of the recoil (since the difference between two mass eigenstates would be proportional to $\Delta m_{ij}^2/(m_Pq_0)\ll1$), by removing the $j$ superscripts from $E_S$ and $E_D$ and fixing
\beq
E_S=\sqrt{m_N^2+\left|\sqrt{q_0^2-m^2}\hat{L}+\vec{p}\p+\vec{p}\p_e\right|^2}\hsp
E_D=\sqrt{m_P^2+\left|-\sqrt{q_0^2-m^2}\hat{L}+\vec{k}\p+\vec{k}\p_e\right|^2}
\eeq
where $m$ is of order $m_j$.  Now assume that the Gaussians on the third line have a much steeper dependence on $q_0$ then the other terms.  By completing the squares, it is possible to rewrite the two Gaussian as one. Let us expand around a certain value of $q_0$, $\hat{q}_0$, which will be chosen in a moment:
\beq
q_0=\hat{q}_0+\delta q_0\hsp
E_S=\hat{E}_S+ v_S \delta q_0\hsp
E_D=\hat{E}_S+ v_D \delta q_0\hsp
\sqrt{q_0^2-m_j^2}=\sqrt{\hat{q}_0^2-m_j^2}+\frac{\delta q_0}{v_j}.
\eeq
Let us also define
\beq
\Delta E_S=\hat{E}_S -\hat{q}_0-p\p_0-p\p_{e,0}\hsp  \Delta E_D=\hat{E}_D +\hat{q}_0-p\p_0-p\p_{e,0}.
\eeq
The exponents of the two Gaussian can be written as
\begin{eqnarray}
&&\sigma^2\left(E^j_S-q_0-p\p_0-p\p_{e,0}\right)^2+\sigma^2\left(E^j_D+q_0-k\p_0-k\p_{e,0}\right)^2=\nonumber \\
&&\sigma^2\left((v_S-1)^2+(v_D+1)^2\right)\left(\delta q +\frac{\Delta E_S(v_S-1)+\Delta E_D(v_D+1)}{(v_S-1)^2+(v_D+1)^2}\right)^2\nonumber \\
&&+\frac{\sigma^2(\Delta E_S(v_D+1)+\Delta E_D(v_S-1))^2}{(v_S-1)^2+(v_D+1)^2}
\end{eqnarray}
and $\hat{q}_0$ can be determined by imposing
\beq
\Delta E_S(v_S-1)+\Delta E_D(v_D+1)=0.
\eeq
Defining
\beq
\sigma_V^2=\sigma^2\left((v_S-1)^2+(v_D+1)^2\right)
\eeq
we can write
\bea
\mathcal{A}_{00}(p\p,p\p_e,k\p,k\p_e)&=&\sum_{j=1}^3 c_j  \left|U_{ej}\right|^2\int d\delta q_0  e^{-i\delta q_0 \left(T-\frac{L}{v_j}\right)}e^{- \sigma_V^2\delta q_0^2}
\eea
for some functions $c_j$ which depend only weakly on $j$. Indeed, the dependence of $c_j$ on the mass eigenstate arises from the following terms
\begin{itemize}
\item $\psi_S\left(\sqrt{q_0^2-m_j^2}\hat{L}+\vec{p}\p+\vec{p}\p_e\right)$ and $\psi_D\left(-\sqrt{q_0^2-m_j^2}\hat{L}+\vec{k}\p+\vec{k}\p_e\right)$. However, in order to see the oscillations the energy width of the source and the energy resolution of the detector should not allow us to discriminate between two different mass eigenstates directly.  Moreover, the fractional dependence is of order $\Delta m^2/q_0^2$, which is of order $10^{-15}$ or less.  So these terms can be neglected.
\item The hadronic currents, $J_{S}^\lambda\left(\vec{p}\p,\sqrt{q_0^2-m_j^2}\hat{L}+\vec{p}\p+\vec{p}\p_e\right)$ and $J_{D}^\rho\left(\vec{k}\p,-\sqrt{q_0^2-m_j^2}\hat{L}+\vec{k}\p+\vec{k}\p_e\right)$. These are related to the cross section for the neutrino production and absorption.  The different mass eigenstates would change the on-shell momenta, but such a difference would be extremely small (for example, in reactor neutrino experiment it would be of the order of $10^{-9}-10^{-10}$ eV), so it would not significantly change the amplitude, and these terms can be considered independent of $j$.
\item The neutrino propagator, {\it i.e.}
\beq
\overline{u}_e(p\p_e)\gamma_\lambda(1-\gamma_5)\left(\slashed{q}|_{\vec{q}=\sqrt{q_0^2-m_j^2}\hat{L}}+m_j\right)\gamma_\rho (1-\gamma_5)v_e(k\p_e).
\eeq
Here it should be noticed that the term proportional to the mass eigenstate is identically zero, %indeed
%\beq
%m_j\overline{u}_e(p\p_e)\gamma_\lambda(1-\gamma_5)\gamma_\rho (1-\gamma_5)v_e(k\p_e)=m_j\overline{u}_e(p\p_e)\gamma_\lambda(1-\gamma_5) (1+\gamma_5)\gamma_\rho v_e(k\p_e)=0
%\eeq
%This means that
only the term proportional to $\slashed{q}$ will remains. Since the direction of the momentum is fixed, the expansion in $m_j$ can be written as
\beq
\simeq \left(1+\frac{\Delta m^2_{ij}}{q_0^2}\right)\overline{u}_e(p\p_e)\gamma_\lambda(1-\gamma_5)\slashed{q}|_{\vec{q}=\sqrt{q_0^2-m_i^2}\hat{L}}\gamma_\rho (1-\gamma_5)v_e(k\p_e)
\eeq
where $m_i$ is one of the neutrino mass eigenvalues that can be chosen arbitrarily.   The fractional $j$ dependence in the first term is again of order $\Delta m^2/q_0^2$.

\end{itemize}
Moreover, all these microscopic terms are not multiplied by any macroscopic quantity, such as the baseline L, unlike the phase, so they can be safely ignored.

We can complete the square and,
% write
%\bea
%\mathcal{A}_{00}(p\p,p\p_e,k\p,k\p_e)&=&\sum_{j=1}^3 c_j  \left|U_{ej}\right|^2
%e^{-\frac{(T-L/v_j)^2}{4\sigma_V^2}} \int d\delta q_0 e^{- \sigma_V^2(\delta q_0+i(T-L/v_j)/(2\sigma_V^2))^2}
%\eea
after performing the last integral and  absorbing the Gaussian normalization factor in the $c_j$'s, we have
\beq
\mathcal{A}_{00}(p\p,p\p_e,k\p,k\p_e)=\sum_{j=1}^3 c_j \left|U_{ej}\right|^2 e^{\frac{(T-L/v_j)^2}{4\sigma_V^2}}
\eeq
or more generally
\beq
\mathcal{A}_{IJ}(p\p,p\p_e,k\p,k\p_e)=\sum_{j=1}^3 c_j \left|U_{ej}\right|^2 e^{\frac{(T+(J-I)\sigma-L/v_j)^2}{4\sigma_V^2}}.
\eeq
We see that the Gaussian for each flavor $j$ is suppressed if the corresponding flavor of neutrino, at velocity $v_j$, did not travel a distance $L/v_j$ in a time within $\sigma$ of $T+(J-I)\sigma$.  For each flavor there will be some expected arrival time, and so some $J-I$ which is not suppressed, but the value of the unsuppressed $J-I$ will depend on the flavor $j$.  In particular, the unsuppressed $J-I$ will be
\beq
(J-I)_j=\frac{L/v_k-T}{\sigma}.
\eeq

\subsection{The Probability}

Now the probability is 
\beq
P(p\p,p\p_e,k\p,k\p_e)=\sum_{I,J=\infty}^{\infty} |\mathcal{A}_{IJ}(p\p,p\p_e,k\p,k\p_e)|^2.
\eeq
The cross terms between different flavors $j_1$ and $j_2$ will be suppressed by Gaussians if the corresponding $(J-I)_j$ differ, which occurs if
\beq
\left| \frac{1}{v_{j_1}}-\frac{1}{v_{j_2}}\right|>>\frac{\sigma}{L}. \label{dcon}
\eeq

In the absence of cross terms, there will be no oscillations.  Therefore we find decoherence in neutrino oscillations when the coherence windows $\sigma$ are sufficiently small or the baseline $L$ is sufficiently large.

In Ref.~\cite{grimus}, the decoherence was lost above Eq.~ (61) when the Gamma term was dropped, making the time interval of creation infinite.  In Ref.~\cite{mcgreevy}, the effect here is present in their Eq.~(27): Notice that the norm of the amplitude is maximized when the creation occurs at time 0, and the detection occurs at a time corresponding to the neutrino propagation time for a given mass eigenstate.  However this time will not agree for different eigenstates, and so the cross terms will be suppressed by the exponential factors in (27).  Similarly in Ref.~\cite{misurapprox} no oscillations were present as there was no constraint on either the creation or the detection times.

%\subsection{Interpreting the Decoherence Bound}

In the ultrarelativistic limit, Eq.~(\ref{dcon}) is very familiar.  The left hand side becomes the velocity difference and $L$ is the propagation time.  Interpreting $\sigma$ as the spatial size of the neutrino wave packets, this is just the old statement that decoherence appears when the wave packets have separated.   Another interpretation, whose consistency with the uncertainty principle merits investigation, is that decoherence occurs when $\sigma$ is less than the inverse of the energy resolution required to observe neutrino oscillations.

%Another interpretation is that, in the ultrarelativistic limit, decoherence only occur if
%\beq
%\sigma<L\Delta v\sim\frac{L \Delta m^2}{E^2 }\sim \frac{L}{E L_{\rm{osc}}}=\frac{N_{\rm{osc}}}{E}
%\eeq
%where $L_{\rm{osc}}$ and $N_{\rm{osc}}$ are the oscillation length and the number of oscillations that have occurred after a distance $L$.  The right hand side is the reciprocal of the energy spacing between oscillation peaks, and so is the reciprocal of the energy resolution needed to observe oscillations.   Thus decoherence occurs if $1/\sigma$ is greater than the required energy resolution.   

%Does the uncertainty principle imply that the uncertainty in the energy is always greater than $1/\sigma$?  If so, Eq.~(\ref{dcon}) would suggest that environmental decoherence is unobservable when the energy resolution is sufficient to observe neutrino oscillations, leaving irreducible decoherence as the leading mechanism.  We will return to this question in future work.

\section{Neutrino Wave Packets}
\label{secModel}

Neutrinos cannot be observed directly but they can be detected via their interactions with other particles. Even if we consider a two-body decay, where the energy of the daughter particles is fixed, any uncertainty on the momentum of the source or detector particle will lead to an uncertainty on the momentum of the neutrino as well. In  Ref.~\cite{mcdonald} it was argued that, if we require a energy resolution sufficient to see the oscillations, the spatial dimension of the detector wave packet would prevent us from observing decoherence. We will show, however, that in the two-body decay the uncertainty on the neutrino momentum will be considerably smaller than that on the source (or detector) particle, due to the on-shell condition, allowing this argument to be evaded and decoherence to emerge.

%In the present section, we will present an example showing that the uncertainty principle does not imply such a bound.  To make as clean an argument as possible, we will make the example as simple as possible.  We will do this by removing both the detection and also by considering a very simple model.  If such a bound is indeed implied by the uncertainty principle, then it will be implied in such simplified settings as well.

\subsection{The Simplified Model}

We have seen that the terms related to the propagator and the hadronic currents in the full 3+1d $V-A$ model contribute only a constant factor to the transition amplitude.  This factor does not affect the oscillations in general nor the decoherence in particular.  For this reason from here thereafter we will consider a simplified model, using only scalar fields in 1+1d, similar to the one used in \cite{Evslin:2019amm,misurapprox}.  In Ref.~\cite{cgl}, it was argued that the physics of both oscillations and entanglement are well captured by a projection to one spatial dimension.

We will consider a model described by a Hamiltonian $H$ which is the sum of the standard kinetic term $H_0$ and an interaction term
\beq
H_I=\int dx :\mathcal{H}_I(x):\hsp
\mathcal{H}_I(x)=\phi_{H}(x)\phi_{L}(x)\left(\nu_1(x)+\nu_2(x)\right)
\eeq
where $\nu_1$ and $\nu_2$ are neutrino mass eigenstates.  Here $\phi_{H(L)}$ represent the heavy (light) source particle and the colons denote the standard normal ordering. The notation $|H,p;0\rangle$ will indicate a state containing a heavy source particle with momentum $p$ and no neutrino, while $| L,k;i,q\rangle$ will contain a light source particle ({\it i.e.} after the decay) with momentum $k$ and a neutrino in a mass eigenstate (with mass $m_i$) with momentum $q$.

In our simplified model, $\nu_1\pm\nu_2$ play the role of the fields for flavor eigenstates $|F_1\rangle\pm|F_2\rangle$.  Our initial state contains only the heavy source particle, that will later decay via the process
\beq
\phi_H(p)\rightarrow\phi_L(p-q)+\nu_i(q)
\eeq
and creates the neutrino in the flavor eigenstate $|F\rangle=\sum_i|F_i\rangle$.  It is
\beq
|I\rangle=\int \frac{\textrm{d}p}{2\pi}G(p-p_0,\sqrt{2}\sigma_I)|H,p;0\rangle\hsp
G(x,\sigma)=\textrm{Exp}\left[-\frac{x^2}{2\sigma^2}\right]
\eeq
where $\sigma_I$ is the dimension of the heavy source wave packet.  This state, like the others we will consider later, is not normalized, since such a normalization factor will not be relevant in the following calculations.
 
%can already be carried out; imposing that $k_0=p_0-q_0$, we have
%\beq \textrm{Exp}\left[-\frac{(p-q-(p_0-q_0))^2}{4\sigma_F^2}-\frac{(q-q_0)^2}{4\sigma_\nu^2}\right]=
%\textrm{Exp}\left[-\frac{(p-p_T)^2}{4\sigma_{F}^2}-\frac{(q-q_0)^2}{4\sigma_\nu^2}\right] \eeq
%where, however, now $p_T$ depends on $q$ as well,
%\beq p_T=q-q_0+p_0 \eeq
We want to compute the transition amplitude
\beq
\mathcal{A}(t)=\langle F|e^{-iHt}|I\rangle=\sum_i\mathcal{A}_i(t) \qquad \mathcal{A}_i(t)=\langle F_i|e^{-iHt}|I\rangle \label{surv}
\eeq
where $|F\rangle=\sum_i|F_i\rangle$ is a flavor eigenstate with fixed momentum.  This amplitude plays the role of the survival amplitude in our crude model, whereas $\langle F_1|-\langle F_2|$ would have led to the appearance amplitude for the other flavor.  At tree-level one finds \cite{ Evslin:2019amm},
\beq
\langle L,p-q;i,q|e^{-iHt}|H,p\rangle \simeq \int \textrm{d}t_1 e^{-i(E_0(p)t_1+E_{1,i}(p-q,q)(t-t_1))}
\eeq
where $t_1$ is the time of creation of the neutrino and
\bea \label{energies}
E_0(p)&=&E_H(p)=\sqrt{p^2+M_H^2}\\
E_{1,i}(p-q,q)&=&E_L(p-q)+E_i(q)=\sqrt{(p-q)^2+M_L^2}+\sqrt{q^2+m_i^2}.\nonumber
\eea
$E_H(p)$ ($M_H$) and $E_L(p-q)$ ($M_L$) are the energies (masses) of the heavy and light source particle, respectively, while $E_i(q)$ ($m_i$) is the energy (mass) of the $i$-th eigenstate of the neutrino mass matrix.

%What we want to do now is to first compute the integrals over the momenta, then integrate over the time. 
Usually in analytic computations these energies are expanded up to the first order in the momentum.  At this order, all of the modes propagate at the same velocity and the wave packet does not spread. 
We will proceed as follows: first we will consider only the first-order approximation, to see how the localization in space of the final state leads to decoherence, then we will see, in one particular case, how the inclusion of the second order terms change the picture.
%We will proceed as following: first, to gain familiarity with this kind of calculations, we will consider only the first order expansion in the momenta, checking if we are able to reproduce the well-known results about decoherence, then we will include the second order expansion as well.

\subsection{First-Order Approximation}

Imposing $E_0(p)=E_{1,i}(p_0-q,q)$ we can solve the on-shell condition for $q$, however its solution will depend (albeit weakly) on $m_i$
\beq
q_{s,i}\simeq \Delta M+\frac{\Delta M (2p_0-\Delta M)}{2M_H} -\frac{m_i^2}{2\Delta M}
\eeq
where $\Delta M$ is the mass difference between the heavy and light source.
We will consider only two neutrino flavors.  To ensure a good convergence of our expansion, let us expand $q$ about $q_0$ which is defined to be the average between the two $q_{s,i}$
\beq
q_0=\frac{q_{s,1}+q_{s,2}}{2}\simeq \Delta M+\frac{\Delta M (2p_0-\Delta M)}{2M_H}-\frac{m_1^2+m_2^2}{4\Delta M}\hsp q_0-q_{s,i}=\frac{\Delta m_{ij}^2}{4\Delta M}\simeq \frac{\Delta m_{ij}^2}{4q_0}
\eeq
where $j\neq i$. We define
\beq \delta p=p-p_0 \qquad \delta q=q-q_0.
\eeq
Expanding $p$ and $q$ about $p_0$ and $q_0$ we can write (\ref{energies}) as
\beq \label{expFirstOrder}
E_0(p)\simeq E_0 +v_I\delta p \qquad E_{1,i}(p-q,q)\simeq E_0+\frac{\Delta m_{ij}^2}{4q_0}+(v_{\nu,i}-v_F)\delta q + v_F\delta p
\eeq
where $v_{\nu,i}$ , $v_I$ and $v_F$ are the group velocities of the neutrino and the heavy and light source particle, respectively and $E_0=E_0(p_0)$. It will be useful to point out that
\beq
v_{\nu,i}=\left.\frac{\partial E_i(q)}{\partial q}\right|_{q=q_0}=\frac{q_0}{2\sqrt{q_0^2+m_i^2}}\simeq 1-\frac{m_i^2}{2q_0}\Rightarrow v_{\nu,i}-v_{\nu,j}\simeq \frac{\Delta m^2_{ij}}{2q_0}.
\eeq

\subsection{Neutrino Wave Packet}
\label{secwave packet}
All the information about the neutrino creation is contained in the time-evolution operator $U=e^{-iHt}$. What is then the size and the shape of the wave packets of the neutrino and the light source particle, after a time $t$ is passed? The state $e^{-iHt}|I\rangle$ contains both the neutrino and the light source particle; the neutrino wave packet in momentum space is defined as
\beq\label{WPexact}
g_\nu(q)=\int \textrm{d}k \langle L,k;i,q|e^{-iHt}|I\rangle \simeq \int \frac{\textrm{d}p}{2\pi} \textrm{d}t_1 e^{-i(E_0(p)t_1+E_{1,i}(p-q,q)(t-t_1))}G(\delta p,\sqrt{2}\sigma_I)
\eeq
where, in the last step, we used the tree-level approximation.  The integral over $k$ was trivial due to momentum conservation.

Expanding $E_0(p)$ and $E_1(p,q)$ up to first order we have
\beq
g_\nu(q)=\int_0^t \textrm{d}t_1 \int \frac{\textrm{d}\delta p}{2\pi} G(\delta p,\sqrt{2}\sigma_I)e^{-i(\rho_{w0,i}+\rho_{wp}\delta p+\rho_{wq,i}\delta q)}
\eeq
where
\begin{eqnarray}
\rho_{w0,i}&=&E_0t+\frac{\Delta m_{ij}^2}{4q_0}(t-t_1) \nonumber \\
\rho_{wp}&=& v_It - (t-t_1) (v_I - v_F)  \nonumber \\
\rho_{wq,i}&=&(t-t_1) (v_{\nu,i} - v_F). \label{rhoWdef}
\end{eqnarray}
Completing the square, the integrals over $p$ and $q$ are now trivial.  We can rewrite $g_\nu(q)$ as
\beq \label{approximateWP}
g_\nu(q)=\frac{\sigma_Ie^{-i\phi_wt}}{\sqrt{\pi}}G(\delta q,\sqrt{2}\sigma_{\nu})I_w(t,q)
\eeq
where (assuming that the neutrino is ultrarelativistic and the source particle, both before and after the decay, is non-relativistic)
\bea
I_w(t,q)&=&\int_0^t\textrm{d}\tilde{t}\textrm{Exp}\left[-\sigma^2_I(v_I-v_F)^2
\left(\tilde{t}-t\frac{v_I}{v_I-v_F}+i\frac{(q-q_{0})(v_{\nu,i}-v_F)}{2\sigma_I^2(v_I-v_F)^2} \right)^2  \right]\label{nusig}\\
\sigma_{\nu,i}&=&\frac{\sigma_I(v_I-v_F)}{v_{\nu,i}-v_F}\simeq \sigma_I(v_I-v_F)\simeq \sigma_Iq_0/M_H\hsp
\phi_w=E_0+(q-q_{0})\frac{v_I(v_{\nu,i}-v_F)}{v_I-v_F}.\nonumber
\eea
%The approximations used to obtain this formula are valid as long as the relevant momenta $p$ and $q$ are sufficiently close to $p_0$ and $q_0$; this is verified when $\sigma_I$ is sufficiently small and $t$ is sufficiently large (otherwise the on-sell and off-shell momenta, respectively, would be too far away from $p_0$ and $q_0$ for the approximation to be valid), as can be seen from Fig. \ref{sigSmall} and \ref{sigBig}. 

\begin{figure} [!tph]
\begin{center}
\includegraphics[width=2.5in,height=1.7in]{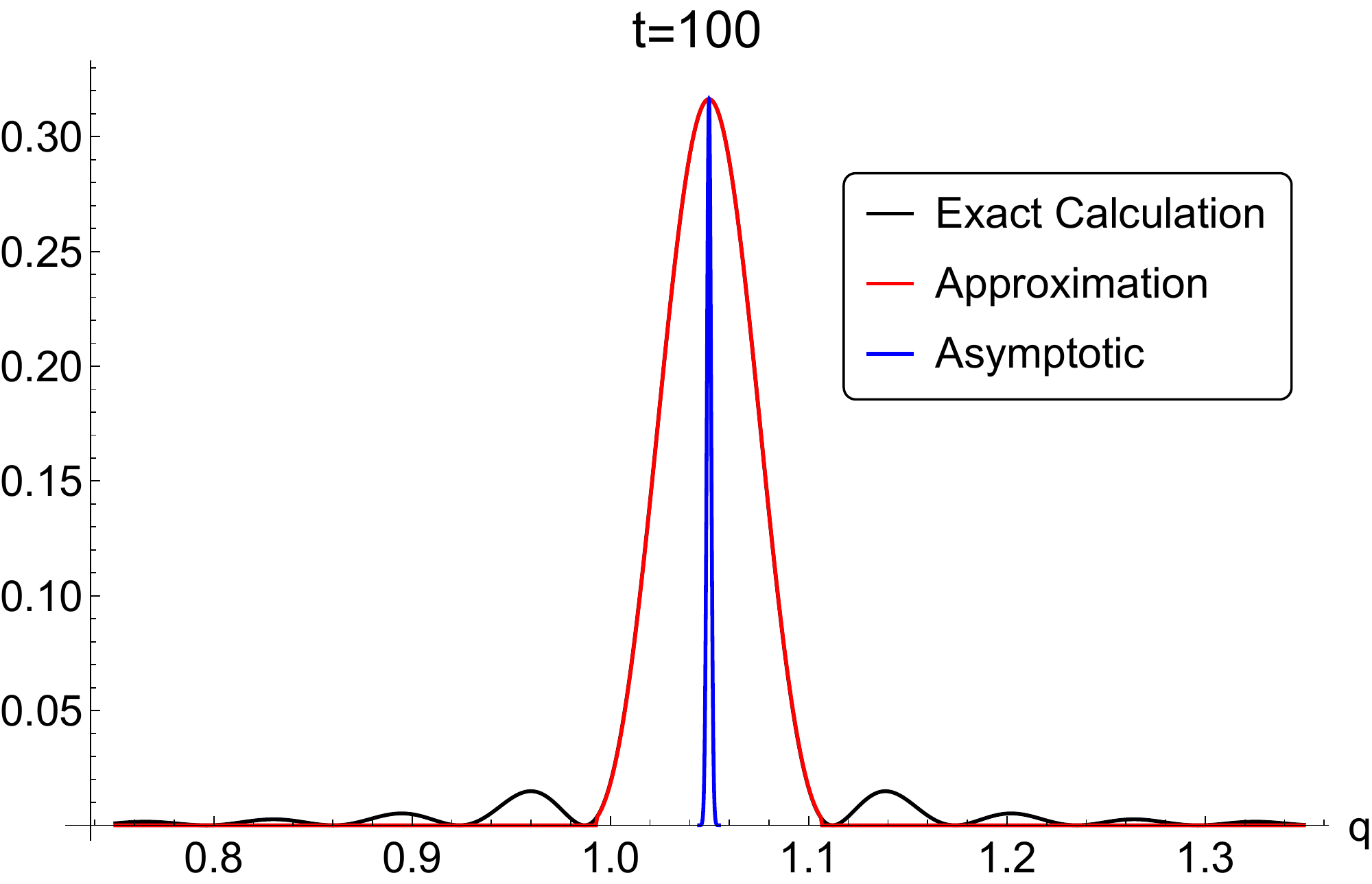}
\includegraphics[width=2.5in,height=1.7in]{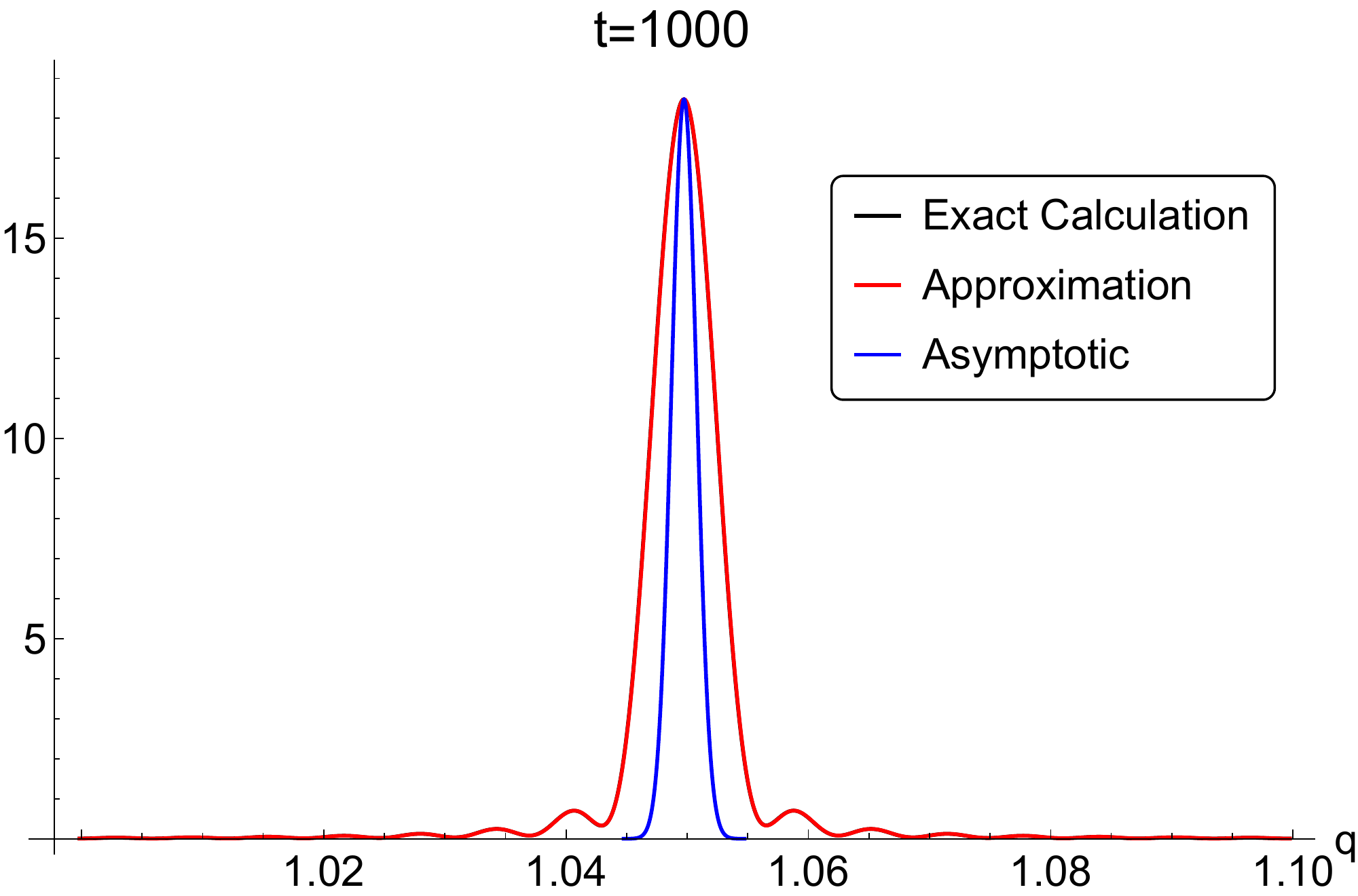}
\includegraphics[width=2.5in,height=1.7in]{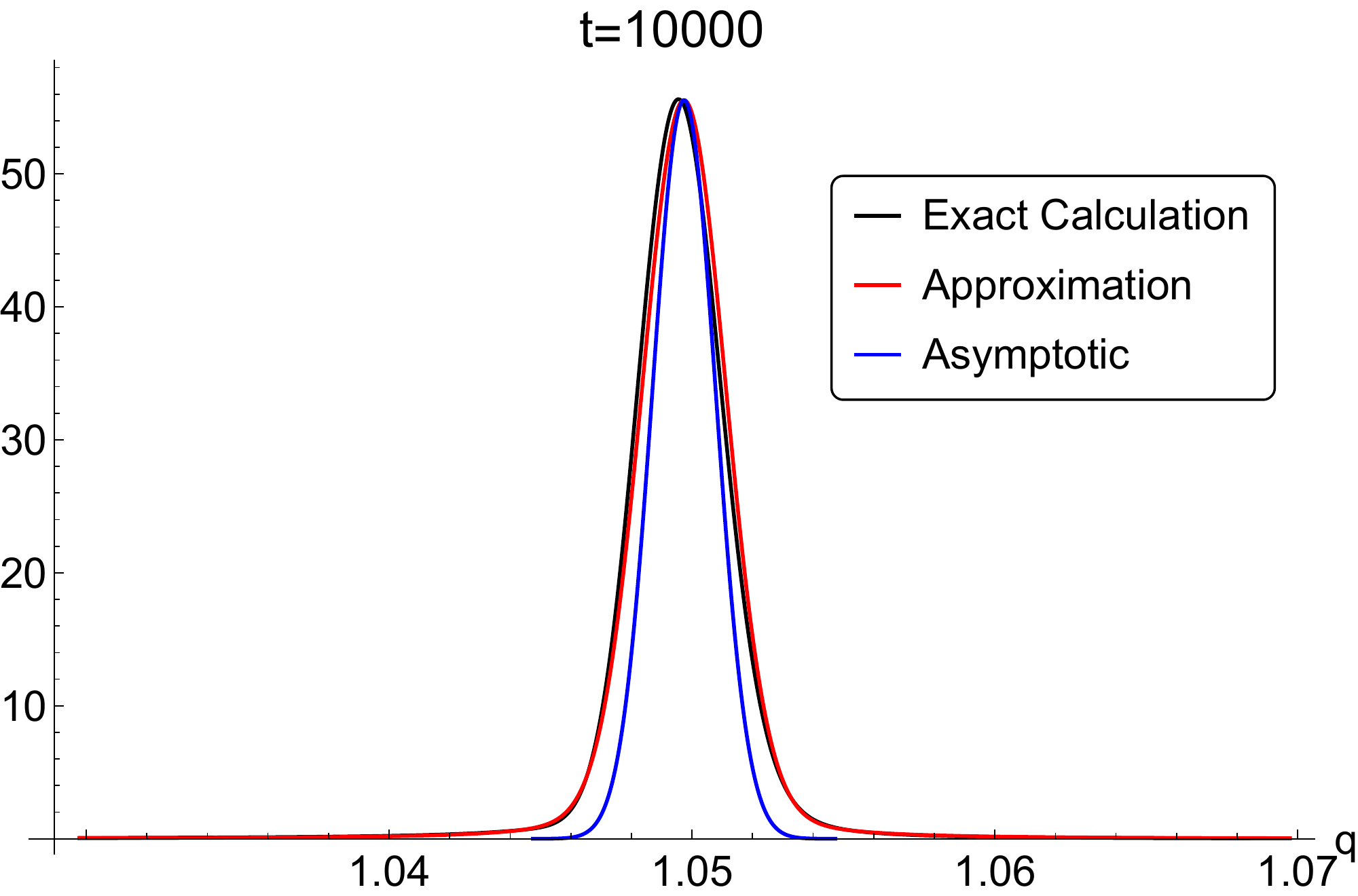}
\includegraphics[width=2.5in,height=1.7in]{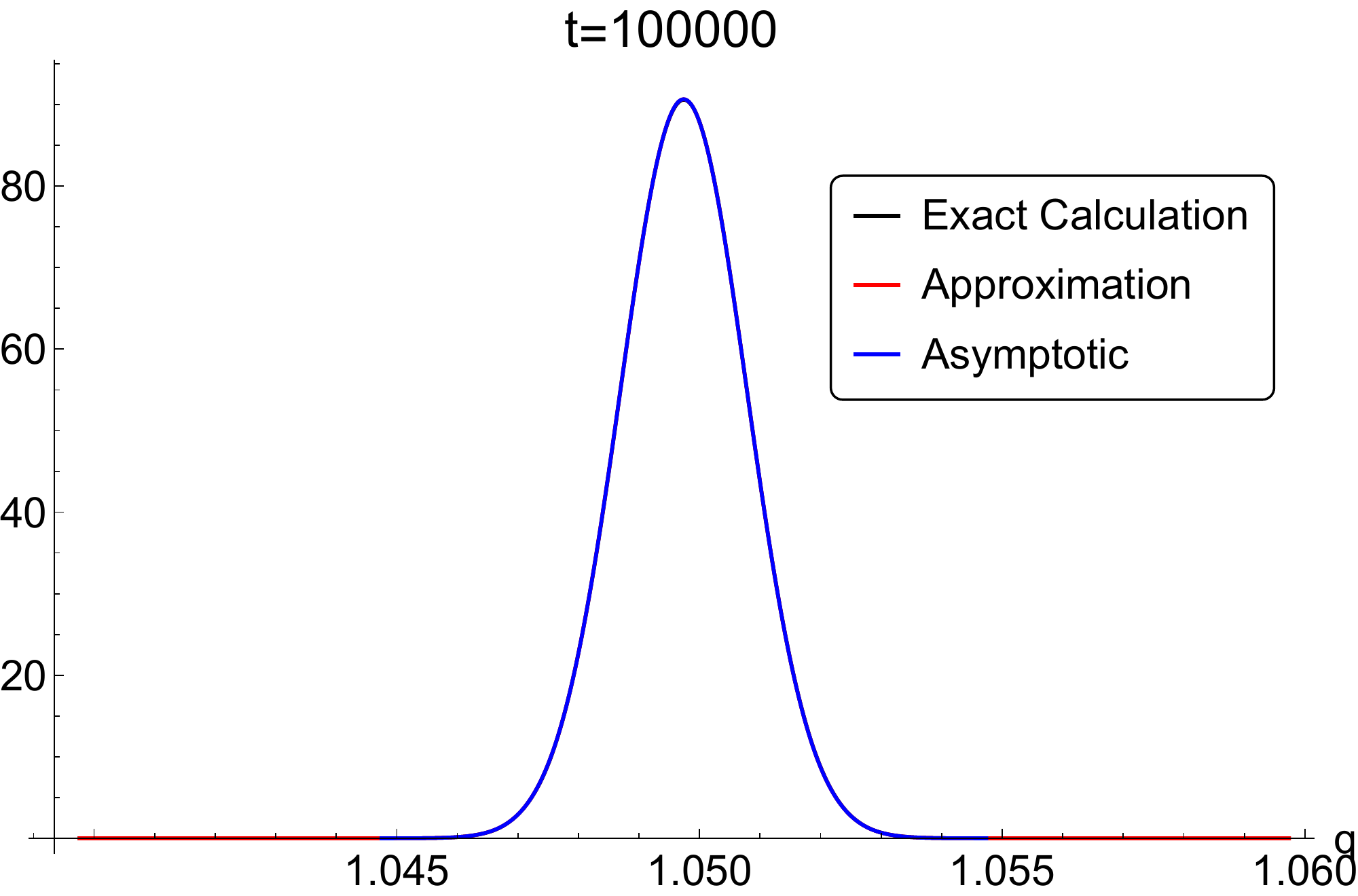}
\caption{The neutrino wave packet in momentum space after a time $t$. The red curve corresponds to our analytical approximation (\ref{approximateWP}), while the black curve is obtained computing Ep. (\ref{WPexact}) via numerical integration.  We considered $\sigma_I=0.01$, $M=10$, $\Delta M=1$, $m_1=0$ and neglected the term proportional to $\Delta m_{ji}^2=0.01$. We can see that when $t\gg t_w\simeq 10^3$ the dimension of the wave packet reaches the asymptotic value, given by the blue curve.}
\label{sigSmall}
\end{center}
\end{figure}

The global phase $\phi_w$ can be neglected, since we are interested in the probability distribution in momentum space, which is proportional to $|g_\nu(q)|^2$. $I_w(t,q)$ represents the contribution of the off-shell momenta; when $t$ is large ($t\gg t_w=1/\sigma_Iv_I$), the q-dependence of $I_w$ can be neglected, and that term is just a multiplicative constant. For large $t$, the wave packet is a Gaussian with $\sigma=\sigma_{\nu}$.  For small $t$, as long as $\sigma_I$ is sufficiently small, our approximation is still valid and in Fig.~\ref{sigSmall} we can see that there is a good agreement between our approximate expression from Eq. (\ref{approximateWP}) and the numerical estimation, however now the $q$ dependence of $I_w$ cannot be ignored anymore and it would significantly increase the width of the wave packet, as well as creating deviations from the Gaussian behavior.

Following the same procedure, we can find the same expression for the wave packet of the light source particle, after the emission of the neutrino.  In particular we have
\beq
\sigma_{F}=\frac{\sigma_I(v_{\nu,i}-v_I)}{v_{\nu,i}-v_F}\simeq\sigma_I \label{slsig}
\eeq
Eq.~(\ref{slsig}) tells us that the wave packet of the final state of the source particle will have more or less the same dimension as the initial one, while the expression for $\sigma_{\nu,i}$ in (\ref{nusig}) shows that the distribution of the neutrino momentum will be much more peaked. The reason for this is that the dimension of these wave packets is determined by imposing energy conservation.  If the mass of the source particle, both before and after the decay, is much larger than the neutrino mass, a change in the initial momentum of the source particle will translate into a much smaller change of the neutrino momentum, hence the spread will be significantly reduced. 

A more intuitive way of obtaining the same result could be the following: let us consider the on-shell condition, which is satisfied for $p=p_0$ and $q=q_0$ (we will ignore, for simplicity, the additional term proportional to $m_i^2$, which is equivalent to assuming that the difference between $q_{s,1}$ and $q_{s,2}$ is very small).  After a slight shift of the momenta we have
\beq
E_0(p_0+\delta p)=E_{1,i}(p_0+\delta p, q_0+\delta q) \Rightarrow \delta p(v_I-v_F)=(v_{\nu,i}-v_F)\delta q \Rightarrow \delta q =\frac{v_I-v_F}{v_{\nu,i}-v_F}\delta p.
\eeq

This resolves the problem suggested in Ref.~\cite{mcdonald}, where the uncertainties in the momenta of the source particle and the neutrino were identified.  We see that the neutrino momentum is more strongly peaked than that of the source particle by a factor of $(v_I-v_F)/(v_{\nu,i}-v_F)$ which is of order the velocity of the nonrelativistic source particle.

\subsection{Coordinate Space Wave Packet}

It will be useful to consider the wave packet in coordinate space as well. It is defined as
\begin{eqnarray}\label{WPexactX}
g_\nu(x)&=&\int \textrm{d}k\textrm{d}q \int\textrm{d}y e^{i(y(k-q)+xq)} \langle L,k;i,q|e^{-iHt}|I\rangle.
\end{eqnarray}
Expanding the energies to first order
\beq
g_\nu(x)=\int \textrm{d}p\textrm{d}q \textrm{d}y \int_0^t\textrm{d}t_1 e^{-i(\rho_{wx0,i}+\rho_{wxq,i}\delta q+\rho_{wxp}\delta p)} G(\delta p,\sqrt{2}\sigma_I)
\eeq
where
\begin{eqnarray}
\rho_{wx0,i}&=&E_0t+\frac{\Delta m_{ij}^2}{4q_0}(t-t_1)-y(p_0-q_0)-xq_0 \nonumber \\
\rho_{wxp}&=& v_It - (t-t_1) (v_I - v_F)-y  \nonumber \\
\rho_{wxq,i}&=&(t-t_1) (v_{\nu,i} - v_F)-(x-y). \label{rhoWdef}
\end{eqnarray}
The integral over $\delta q$ (which was not present when we computed the wave packet in momentum space) gives us a $\delta ((t-t_1) (v_{\nu,i} - v_F)-(x-y))$, which gets rid of the integral over $y$; the integral over $\delta p$ is now a simple Gaussian. We have (up to a global phase and an irrelevant normalization factor)
\beq
g_\nu(x)\propto \int_0^t \textrm{d}t_1 G( v_It_1 + v_{\nu,i}(t-t_1)-x, 1/(\sqrt{2}\sigma_I) ).
\eeq
We can see that, as long as the peak of the Gaussian is within the integration domain ({\it i.e.} it is possible for the neutrinos created in the time window we are considering to arrive at $x$), this integral will be constant and will not depend on $x$.  This means that the dimension of the wave packet in coordinate space will grow linearly in $t$, while in momentum space it will reach an asymptote. Thus, even if the neutrino wave packet in momentum space is described by a Gaussian, its dimension in coordinate space could be significantly larger than $1/\sigma_q$. 

This growth of the wave packet is not in contradiction with the uncertainty principle, which provides only a lower bound on the product of these uncertainties.  Note that it is not to be confused with the spreading of the wave packet, studied below, which results from the spread in the velocity of the neutrinos.  Rather this growth is caused by the continuous neutrino production.

\section{Transition Amplitude}
\label{secAmp}

In Sec.~\ref{secGrimusGen} we showed that decoherence results if the quantum numbers of the environment determine both the neutrino's emission and absorption times.  This was achieved by simply postulating an amplitude which determines the final state of the environment as a function of the emission and absorption time.  For the sake of generality, we did not consider a specific interaction which generates that amplitude.  This generality of course comes at a cost, one may wonder if any such interaction exists.  

\subsection{Gaussian Wave Packets}

This motivates us, in the present section, to present an independent and more conventional demonstration that if the final state determines the emission and absorption times, decoherence results.  We rely on the observation of Ref.~\cite{Giunti:2002xg} that environmental interactions have the effect of measuring the position of the final state of the unobserved particle $\phi_L$.  Therefore, in this section, we will consider the amplitude (\ref{surv}) in which the final state $|F\rangle$ is no longer a momentum eigenstate, but now a state in which both the neutrino and also $\phi_L$ are spatially localized.  The spatial localization of the former is our crude approximation of measurement, while that of the later is our proxy for environmental interactions.

By measuring the locations of the final state particles at a fixed time, the production point is determined and so, as we do not consider absorption here, the arguments above suggest that decoherence should result.  We will see that this is indeed the case, providing further evidence for our general description of decoherence.

%We have seen in \cite{misurapprox} that, if the neutrino is created in the vacuum ({\it i.e.} no environmental interactions) and if the final states are plane waves there is no decoherence due to the separation of the mass eigenstates. This happens because, under these conditions, the neutrino production is not observed and the time-evolved state is a superposition of all the possible creation times: as was pointed out in \cite{as10}, decoherence is related to the localization in space (or, equivalently, in time) of the neutrino production and detection. We have seen in Sec. \ref{secGrimusAmp} that such a localization can be obtained by turning on the interactions only for a limited time windows; our main goal in this section will be to show that the same result can be achieved if both the initial and final state are localized. We will consider the final state

More specifically, we will consider the amplitude (\ref{surv}) with respect to the final state
\beq\label{final}
|F\rangle=\sum_i|F_i\rangle=\sum_i\int \frac{\textrm{d}q\textrm{d}k}{(2\pi)^2}e^{-ikL_F}e^{-iqL_\nu}G(q-q_0,\sqrt{2}\sigma_\nu)G(k-k_0,\sqrt{2}\sigma_F)|L,p-q;i,q\rangle
\eeq
where $\sigma_{\nu(F)}$ and $L_{\nu(F)}$ are, respectively, the dimension of the wave packet  and the position at time $t$ of the neutrino (light source particle). It should be noticed that the conservation of momentum will impose $k=p-q$, so the integral over $k$ will be trivial. 

It is convenient to write the product of the Gaussians as
\beq\label{prdGaussian}
G(\delta p,\sqrt{2}\sigma_I)G(\delta q,\sqrt{2}\sigma_\nu)G(\delta p-\delta q,\sqrt{2}\sigma_F)= G(\delta p -\alpha \delta q,\sigma_p)G(\delta q, \sigma_q)
\eeq
where
\beq
\frac{2}{\sigma_p^2}=\frac{1}{\sigma_I^2}+\frac{1}{\sigma_F^2}\hsp
\frac{2}{\sigma_q^2}=\frac{1}{\sigma_I^2+\sigma_F^2}+\frac{1}{\sigma_\nu^2} \hsp
\alpha=\frac{\sigma_p^2}{2\sigma_F^2}. \label{sigDef}
\eeq
Following the same procedure used in the previous section, the linear approximation allows us to write the transition amplitude as
\beq\label{defAmplitude}
\mathcal{A}_i(t)=\int_0^t \textrm{d}t_1 \int \frac{\textrm{d}\delta p\textrm{d}\delta q}{(2\pi)^2} G(\delta p -\alpha \delta q,\sigma_p)G(\delta q, \sigma_q)e^{-i(\rho_{0,i}+\rho_p(\delta p-\alpha\delta q) +\rho_q\delta q )}
\eeq
where
\begin{eqnarray}\label{defRhoEqArray}
\rho_{0,i}&=&E_0t+\frac{\Delta m_{ij}^2}{4q_0}(t-t_1)-p_0L_F-q_0(L_\nu-L_F) \nonumber \\
\rho_{p}&=&( v_It-L_F)-(t-t_1) (v_I - v_F) \nonumber \\
\rho_q&=&(t-t_1) (v_{\nu,i} - v_F) - (L_\nu - L_F)+\alpha\rho_p.
\end{eqnarray}
Completing the square, and integrating first over $p$ and then over $q$, we  can write the transition amplitude
\beq
\mathcal{A}_i(t)=e^{-i\theta+i\rho_{T,i}}e^{-\delta_i^2/2}I
\eeq
where $\theta$ is a global phase, that can be neglected since it does not depend on the mass eigenstate $i$,
\begin{eqnarray}
I&=&\int_0^t\textrm{d}\tilde{t}e^{-(a^2+c_i^2)(\tilde{t}+\frac{ab+c_id+ie_i}{a^2+c_i^2})^2/2}\label{intOverT}\nonumber \\
\rho_{T,i}&=&\frac{(ab+c_id)e_i}{a^2+c_i^2} \hsp \delta_i^2=\frac{(bc_i-ad)^2+e_i^2}{a^2+c_i^2}\label{defRho} \nonumber\\
a&=&-\sigma_p(v_I-v_F) \hsp b=(v_It-L_F)\sigma_p \nonumber \\
c_i&=&\sigma_q(v_{\nu,i}-v_F+\alpha(v_F-v_I)) \hsp d=-(L_\nu-L_F+\alpha(L_F-v_It))\sigma_q \nonumber \\
e_i&=&\frac{\Delta m_{ij}^2}{4q_0} \hsp \theta=E_0t-p_0L_F-q_0(L_\nu-L_F)\label{defABC}
\end{eqnarray}
and $\tilde{t}=t-t_1$. 

$L_\nu$ and $L_F$ are parameters that can be arbitrarily chosen and correspond to the position of the neutrino and the light source particle at time $t$.  Note that, at $t=0$, the heavy source particle is located in the origin of our reference system. 

These quantities are intrinsically related to the localization of the neutrino production: classically, if we know all the masses and the momenta involved, the position of the light source particle at time $t$ would determine the moment when the neutrino was created. However, in QFT, particles do not occupy a precise position in space, so instead of "position", $L_F$ would only determine the "region" where the neutrino was created, let's call it $\mathcal{R}_F$.  Similarly $L_\nu$ determines a different creation region for each mass eigenstate, $\mathcal{R}_{\nu,i}$.  If all the $\mathcal{R}_{\nu,i}$'s have a significant overlap with $\mathcal{R}_F$, no mass eigenstate is suppressed and we would see the oscillations, otherwise we would have decoherence.

%Indeed, as we have seen in Sec. \ref{secwave packet}, in the time-evolved initial state we have two particles, the neutrino and the light source, which at large $t$ are described by Gaussian wave packets. It is however a superposition of states, given by the coherent integral over all the possible times at which the neutrino can be created, so the space localization will be lost; we are now projecting such a state into a final state which, instead is localized in space, where the two particles are still described by Gaussian wave packet, centered in $L_f$ and $L\nu$; as can be seen in Eq. (\ref{intOverT}), this will constrain the values of $\tilde{t}$ that will give a non-negligible contribution to the transition amplitude, with a $\sigma_t\simeq 1/(\sigma_p^2(v_I-v_F)^2+\sigma_q^2)$, which depends on the precision with which the heavy source particle and the final states are localized. 
Let us write 
\beq \label{defLnu}
L_\nu=v_I (t-t_p) + v_{\nu,0}t_p \qquad L_F=L_{F,0}+\delta L \qquad L_{F,0}=v_I (t-t_p) + v_F t_p 
\eeq
where $t_p$ and $v_{\nu,0}$ are two arbitrary parameters that must be introduced by hand.  They can be identified as the neutrino propagation time and the velocity of the neutrino in the classical limit, respectively. For a given value of the set $\{L_F,L_\nu,i,v_{\nu,0}\}$, there will always be a value of $\tilde{t}$ for which either the light source particle will be at $L_F$ or the neutrino will be at $L_\nu$ (we have to consider the additional constraint $0<\tilde{t}<t$, but that's not relevant for this discussion), however  these conditions can not necessarily be satisfied at the same time. This happens, however, when $\delta L=0$ and $v_{\nu,0}=v_{\nu,i}$, which corresponds to the case where the neutrino and the light source particle, at time $t$, are both in the position that would be expected classically if they were created at the time $t-t_p$ and the velocity of the mass eigenstate we are considering is exactly equal to $v_{\nu,0}$.  It is easy to verify that, in this case $\delta^2_i$ from Eq. (\ref{defRho}) vanishes, however this cannot be done for more than one mass eigenstate at the time, which means that the other transition amplitudes can be suppressed, leading to decoherence.

The transition probability $P(t)$ (neglecting the multiplicative factor $I$ which, as we pointed out, can be considered constant and independent of $i$) reads
\begin{eqnarray}
P(t)&=&\left|\sum_i \mathcal{A}_i\right|^2=|\mathcal{A}_1|^2+|\mathcal{A}_2|^2+\mathcal{A}_1^*\mathcal{A}_2+\mathcal{A}_2^*\mathcal{A}_1\propto e^{-\delta_1^2}+e^{-\delta_2^2}+2e^{-(\delta_1^2+\delta_2^2)/2}\textrm{Cos}(\rho_{T,2}-\rho_{T,1})\nonumber\\
&=&\left(e^{-\delta_1^2/2}+e^{-\delta_2^2/2}\right)^2-4e^{-(\delta_1^2+\delta_2^2)/2}\textrm{Sin}^2\left(\frac{\rho_{T,1}-\rho_{T,2}}{2}\right). \label{probGen}
\end{eqnarray}

The terms $\delta^2_i$ and $\rho_{T,i}$, however, do not have a simple form and it would be difficult to compare Eq. (\ref{probGen}) with other results available in literature. It is possible to simplify these expressions, however we must be careful.  For example, one may be tempted to consider $v_F\simeq v_I\simeq 0$, since usually the source particle is very heavy and non-relativistic.  However in this case, assuming $\delta L=0$, one obtains $\delta^2_i=0$. This is not surprising, since we have seen previously that we expect the dimension of the neutrino wave packet to be proportional to $\sigma_I(v_I-v_F)$.  If we consider $v_F\simeq v_I\simeq 0$, this is equivalent to considering the neutrino to be completely delocalized, so the only suppression to the transition probability would come from the displacement of the final state of the source particle and not from the different velocities of the neutrino mass eigenstates.

For this reason, we will consider $\sigma_\nu \simeq \sigma_I(v_I-v_F)$ and $\sigma_F\simeq \sigma_I$. Assuming $v_I, v_F\ll 1$ and considering only the leading order, Eq. (\ref{sigDef}) now reads
\beq
\sigma_p\simeq \sigma_I \qquad \sigma_q\simeq \sqrt{2}\sigma_I(v_I-v_F) \qquad \alpha\simeq \frac{1}{2}.
\eeq
It is easy now to compute the denominator that appears in $\delta_i^2$ and $\rho_{T,i}$ at leading order
\beq
a^2+c_i^2\simeq 3\sigma_I^2(v_I-v_F)^2\simeq 3\sigma_\nu^2.
\eeq
It should be noticed that there should be some additional terms, proportional to $v_{\nu,i}-1$.   However, since we already know that the $\delta_i^2$'s depend on the difference between the velocities of the mass eigenstates, and $\rho_{T,i}$ is proportional to $\Delta m_{ij}^2$, these terms would only contribute at the next-to-leading order, and they can be safely neglected.

Let us take a look at $\delta_i^2$.  We can safely neglect the term proportional to $e_i^2$, since it goes like $(\Delta m_{21}^2)^2$.  Taking into account Eq. (\ref{defLnu}), we have
\beq
\delta_i^2=\frac{2}{3}\left(t_p\sigma_\nu(v_{\nu,0}-v_{\nu,i})+\sigma_I\delta L(v_{\nu,i}-v_I)    \right)^2 \simeq 
\frac{2}{3}\left(t_p\sigma_\nu(v_{\nu,0}-v_{\nu,i})+\sigma_I\delta L   \right)^2
\eeq
where $v_{\nu,0}$ is the parameter introduced in Eq. (\ref{defLnu}) as the classical velocity of the neutrino. This term is the one that controls the decoherence. The term proportional to $t_p$ is the one related to decoherence itself, {\it i.e.} the dampening of the oscillations due to the separation of the wave packets.  As expected, it depends on the dimension of the neutrino wave packet (which is considerably smaller, in momentum space, with respect to the one of the source particle). We can also see that, if $v_{\nu,i}\neq v_{\nu,0}$, for each $t_p$ there is a specific value of $\delta L$ that makes $\delta_i^2$ vanish.  This corresponds to the case where the shift due to the difference between the velocities and the shift with respect the classical position compensate each other.

Considering only the leading order, $\rho_{T,i}$ is
\beq
\rho_{T,1}=\frac{\Delta m^2}{4q_0}\left(t_p-\frac{\delta L}{3(v_I-v_F)}\right) \qquad \rho_{T,2}=-\frac{\Delta m^2}{4q_0}\left(t_p-\frac{\delta L}{3(v_I-v_F)}\right)
\eeq
where $\Delta m^2=\Delta m_{21}^2$.

\subsubsection{$\delta L_W=0$}
In order to simplify the calculations, let us choose $v_{\nu,0}=v_{\nu,1}$. If $\delta L_{W}=0$, then we have
\beq
\delta_1^2=0 \qquad \delta_2^2=\frac{2}{3}t_p^2\sigma_\nu^2\left(v_{\nu,1}-v_{\nu,2}\right)^2\simeq\frac{2}{3} \left(\frac{L\Delta m^2}{2E^2\sigma_{x,\nu}}\right)^2
\eeq
where $E=|q_0|$ is the neutrino energy, $L\simeq t_p$ is the baseline for neutrino oscillations, $\sigma_{x,\nu}=1/\sigma_{\nu}$ and, in the last step, we have used the approximation
\beq
v_{\nu,2}-v_{\nu,1}\simeq \frac{\Delta m^2}{2E^2}.
\eeq
We define $L_{coh}$ as
\beq
L_{coh}=\frac{2E^2\sigma_{x,\nu}}{\Delta m^2}.
\eeq
Neglecting the multiplicative factor of 4, that contributes to the total rate but not to the oscillation probability, (\ref{probGen}) can be written as
\beq
P(t)\propto \left(\frac{1+e^{-2L^2/6L_{coh}^2}}{2}\right)^2-\textrm{Sin}^2\left(\frac{\Delta m^2 L}{4E}\right)e^{-2L^2/(6L_{coh}^2)}.
\eeq
If the final state is localized, then, we will have decoherence, which would not be present if our final state $|F\rangle$ were a momentum eigenstate as in Ref.~\cite{misurapprox}.   In other words, decoherence occurs because the environment has constrained the position of $\phi_L$.  The reason for this can be found in Eq.~(\ref{intOverT}), from which we can see that the localization of the final state implies a localization of the production time of the neutrino as well, similarly to the direct environmental constraint on the production time used in Sec. \ref{secGrimus} or in Ref.~\cite{grimus}, where the time constraints are introduced directly in the Hamiltonian, by turning on the interactions only during a given time window. 

We can see that the limit $\tau\rightarrow\infty$, considered in \cite{grimus}, is equivalent to the limit
\beq
a^2+b^2\simeq \sigma_p^2(v_I-v_F)^2+\sigma_q^2\rightarrow 0
\eeq
which would eliminate decoherence in the simple model we considered as well. Decoherence could still emerge, however, if the time of creation of the neutrino is determined more precisely: this is equivalent to the localization in space of the final state of the source particle $\phi_L$ by environmental interactions.

\subsubsection{$\delta L \neq 0$}
Let us return to the case of an experiment in a vacuum.  Imagine that the final state of $\phi_L$ is not entangled to the environment.  Then one needs integrate over all the possible values of its position $\delta L$. We have
\bea
\delta_1^2&=&\frac{2\delta L^2\sigma_I^2}{3} \hsp \delta_2^2=\frac{2}{3}\left( \frac{L}{L_{coh}}+\delta L\sigma_I\right)^2\nonumber\\
\frac{\rho_{T,1}-\rho_{T,2}}{2}&=&\frac{\Delta m^2}{4E}\left(L-\frac{\delta L}{3(v_I-v_F)}\right).
\eea
The transition probability as a function of $\delta L$ reads
\begin{eqnarray}
P(t,\delta L)&=&\left(\frac{e^{-2\delta L^2\sigma_I^2/6}+e^{-2(L/L_{coh}+\delta L\sigma_I)^2/6}}{2}\right)^2  \\
&-&\textrm{Sin}^2\left(\frac{\Delta m^2}{4E}\left(L-\frac{\delta L}{3(v_I-v_F)}\right)\right)
\textrm{Exp}\left[-\frac{2}{6}\left(\delta L^2\sigma_I^2+\left(\frac{L}{L_{coh}}+\delta L\sigma_I \right)^2\right)\right].\nonumber 
\end{eqnarray}
All the possible positions of the final state of the source particle should be summed incoherently, since they all correspond to a different final state. We have
\begin{eqnarray}
P(t)&=&\int\textrm{d}(\delta L_W) P(t,\delta L_W)=\nonumber \\
&&\sqrt{\frac{3\pi}{2\sigma_I^2}}\left( \frac{1+ e^{-\frac{L^2}{6L_{coh}^2}-\frac{(\Delta m^2\sigma_{\nu,x})^2}{6(4E)^2}}}{2} -e^{-\frac{L^2}{6L_{coh}^2}-\frac{(\Delta m^2\sigma_{\nu,x})^2}{6(4E)^2}}
\textrm{Sin}^2\left[ \frac{\Delta m^2L}{4E}\left(1-\frac{\sigma_{\nu,q}}{6L_{coh}} \right) \right]  \right)\nonumber \\
\end{eqnarray}
where $\sigma_{\nu,x}=1/\sigma_q$. By integrating over $\delta L$ we are taking into account all the possible production regions which would allow the neutrino to be detected in a certain point.  This in equivalent to an uncertainty on the baseline, related to the finite dimension of the wave packets, as is found, for example in Refs. \cite{Giunti:2002xg,Giunti:1997sk}. In principle this would depend on the dimensions of all the wave packet involved.  However, since $\sigma_\nu\ll \sigma_I, \sigma_F$, the spatial dimension of the neutrino wave packet is considerably larger that the ones of the source particle (before and after the neutrino emission), so this term is dominated by $\sigma_{\nu,x}$.  

There is still some decoherence when only the neutrino's position is measured.  This is to be expected as the neutrino position also contains information regarding the production time.  We suspect that, including a detector as in Sec.~\ref{secGrimusGen}, the sensitivity of the final state position to the production time would be reduced.  However, we will leave a methodical study of the irreducible decoherence expected in this case to a companion paper.

\subsection{Second Order Expansion}
It is well known that the size of a Gaussian wave packet increases with time, since different $p$-components of the state will travel at different velocities.  Such wave packet spreading, however, is not present in the first order expansion in energy. We will consider now the expansion of the neutrino energy up to the second order, while keeping only the terms up to the first order for the source particles (light and heavy). While $E_0(p)$ remains the same, $E_{1,i}(p-q,q)$ from Eq. (\ref{expFirstOrder}) now reads
\beq
E_{1,i}(p-q,q)\simeq E_0+\frac{\Delta m_{ij}^2}{4q_0}+(v_{\nu,i}-v_F)\delta q + v_F\delta p+v'\delta q^2
\eeq
where
\beq
v'=\frac{1}{2}\frac{\partial E_{\nu}(q)}{\partial q}\simeq \frac{m_i^2}{2E^3}.
\eeq
Since the additional term is proportional to $\delta q^2$, we can incorporate it in $G(\delta q, \sqrt{2}\sigma_q)$.  This is equivalent to replacing $\sigma_\nu$ with $\tilde{\sigma}_\nu$
\beq
\sigma_\nu \rightarrow \tilde{\sigma}_\nu=\sigma_\nu\Sigma(\tilde{t}) \qquad \Sigma^2(\tilde{t})=\frac{1-i\sigma_q^2v'\tilde{t}}{1+(\sigma_q^2v'\tilde{t})^2}
\eeq
where $\sigma_p$ and $\sigma_q$ are defined as in Eq. (\ref{sigDef}).  If we assume again $\sigma_F\simeq \sigma_I$ and $\sigma_\nu\simeq \sigma_I(v_I-v_F)$, we obtain
\beq
\sigma_p\simeq \sigma_I \qquad \tilde{\sigma}_q\simeq \sqrt{2}\sigma_\nu\Sigma(\tilde{t}).
\eeq
We can now follow the same procedure used in the previous section, taking care to substitute $\tilde{\sigma}_q$ for $\sigma_q$ . The integration over $\tilde{t}$ is not trivial, however, since now $\tilde{\sigma}_q$ depends on $\tilde{t}$ as well.  If we ignore $\Sigma(\tilde{t})$ and if the wave packets are sufficiently peaked, the Gaussian over $\tilde{t}$ can be approximated by $\delta(\tilde{t}-t_p)$.  Moreover for large $\tilde{t}$, $\Sigma(\tilde{t})\rightarrow 0$, which does not affect the validity of such an approximation. We obtain then an expression similar to Eq. (\ref{probGen}), where however now we have
\begin{eqnarray}
a^2+c_i^2&=&(1+2\Sigma_p^2)\sigma_\nu \nonumber \\
\delta_i^2&=&\frac{2\Sigma_p^2}{1+2\Sigma_p^2}\left(\sigma_\nu t_p(v_{\nu,0}-v_{\nu,i})+\sigma_I\delta L  \right)^2 \nonumber \\
\rho_{T,i}&=&-\frac{\Delta m_{ij}^2}{4q_0}\left(t_p-\frac{\delta L}{1+2\Sigma_p^2}\right)
\end{eqnarray}
where $\Sigma_p=\Sigma(t_p)$. It should be noticed, however, that since $\Sigma_p^2$ is complex, now $\delta_i^2$ will have an imaginary part as well, that will contribute to the oscillatory behavior. It is convenient then to define
\begin{eqnarray}
\delta_{s,i}^2&=&\textrm{Re}[\delta_i^2] \nonumber \\
\rho_{Ts,i}&=&\rho_{T,i}-\textrm{Im}[\delta_i^2]/2.
\end{eqnarray}
Let us define
\[
L_{sp}=\frac{1}{\sigma_q^2v'}.
\]
We can see that, if $L\gg L_{sp}$, then $\Sigma_p\rightarrow 0$, however $\textrm{Re}(\Sigma_p^2 )L^2\rightarrow \textrm{const}$. If we consider $\delta L=0$, and using the usual approximations employed in this paper, we have
\[
\delta_{s,1}^2=0 \qquad \delta_{s,2}^2=\frac{L_{sp}^2}{L_{coh}^2}
\]
which means that, when $L\geq L_{sp}$, the decoherence ``saturates" and the oscillations are not dampened anymore. Moreover, if we consider the oscillation term, we have
\[
\rho_{Ts,2}-\rho_{Ts,1}=\rho_{T,2}-\rho_{T,1}-\textrm{Im}[\delta_{2,i}]=\left(\frac{\Delta m^2}{4E}+\frac{L_{sp}}{L_{coh}^2}\right)L
\]
which means that the spread of the wave packet can affect the oscillation length as well.

\section{Remarks}

We have shown that the details of the $V-A$ model do not affect the presence of decoherence.  This allowed us to use a simplified model in 1+1d and containing only scalar fields. Using this model we have shown that, if the initial and final states are localized, for example if the neutrino is measured and the daughter particles are entangled with the environment, then the neutrino production time is limited by constraints similar to the ones introduced in  Sec.~\ref{secGrimusGen}, where we saw that they lead to decoherence. %(while, in Ref. \cite{misurapprox}, where the final states were plane waves, no decoherence was present).  

If the source particle is not in a momentum eigenstate, there will be an uncertainty on the neutrino momentum as well.  However the spread of the neutrino wave packet will be considerably smaller than that of the source particle: it will be rescaled by a factor $v_I-v_F$.  If the source is non-relativistic (which would be the case, for example, for rector neutrinos), this factor is very small. This provides a counterexample for one step in the argument of Ref. \cite{mcdonald}, where it was argued that decoherence can never be observed as it appears only when the detector's energy resolution is too poor to see the oscillations.

Often in analytical calculations the energies of the particles are expanded only up to the first order in the momenta.  A consequence of such an approximation is that each momentum component of the wave packet would travel at the same velocity. We have shown that, if the second order of the expansion is taken into account as well, the spread of the wave packet will prevent decoherence from completely canceling the oscillations, which would be damped but still present.

The arguments here have been rather formal.  How can they be converted into a quantitative estimate of neutrino decoherence?

In Sec.~\ref{secGrimusGen} we computed the decoherence resulting from a given time window.  This can straightforwardly be generalized to a given space-time window for the production, and for the detection.   One simply adds an environmental quantum number for each dimension and introduces an amplitude similar to (\ref{pamp}) and (\ref{damp}) for each quantum number.  Thus, if there is no preferred spatial direction, one expects an oscillated spectrum which depends on four decoherence parameters, the temporal and spatial window size at the production and the absorption point.  Next, for any given environmental coupling, one can calculate these four parameters by calculating the corresponding amplitude.  For this purpose, the production process and absorption processes may be considered separately, each yielding the corresponding two parameters. 

In the literature, similar windows have been constructed.  In the external wave packet approach \cite{beuthe}, the allowed space-time window is constructed as the intersection of a set of rigid wave packets.  The shape of this window is entirely determined by the kinematics of the particles directly involved in neutrino production and absorption and is independent of the kinematics of the particles in the environment.  The presence of the environment is only encoded in the overall sizes of the wave packets.  In some studies \cite{kw98}, on the other hand, a time window is used with no spatial constraint.   These two approaches lead to specific constraints on the four parameters above, relating the temporal and spatial size of each window.  However, we claim that the four parameters above are instead determined by the microphysics of the environmental interactions which need not obey such constraints.   Thus we expect that such a full, microscopic approach will lead to a quantitative disagreement with previous studies.

\section* {Acknowledgement}

\noindent
JE is supported by the CAS Key Research Program of Frontier Sciences grant QYZDY-SSW-SLH006 and the NSFC MianShang grants 11875296 and 11675223. EC is supported by the Chinese Academy of Sciences Presidents International Fellowship Initiative Grant No. 2020FYM002. JE and EC also thank the Recruitment Program of High-end Foreign Experts for support.

\end{document}